\documentclass[12pt]{article}
\usepackage{float}
\usepackage{epsfig}
\usepackage{graphicx}
\usepackage{amssymb,amsmath}
\usepackage{physics}
\usepackage{hyperref}
\usepackage[nosort]{cite}

\hypersetup{
    colorlinks=true,
    linkcolor=blue,
    citecolor=red,
}

\usepackage{array}

\begin{document}

\begin{center}
\Large{\Large \bf Star-exponential for Fermi systems and the Feynman-Kac formula}
\vspace{0.7cm}

{\normalsize J. Berra-Montiel$^{a,}$\footnote{e-mail address: {\tt
		jasel.berra@uaslp.mx}}, H. Garc\'{\i}a-Compe\'an$^{b,}$\footnote{e-mail address: {\tt
		hugo.compean@cinvestav.mx}}, A. Kafuri$^{b,}$\footnote{e-mail address: {\tt
anuar.kafuri@cinvestav.mx}}, A. Molgado$^{a,}$\footnote{e-mail
address: {\tt alberto.molgado@uaslp.mx}}} \\

\vspace{0.5cm}

{\small \em $^a$Facultad de Ciencias,  Universidad Aut\'onoma de San Luis Potos\'{\i}\\ 
Campus Pedregal, Av. Parque Chapultepec 1610, Col. Privadas del Pedregal,
San Luis Potos\'{\i}, SLP, 78217, Mexico}\\

\vskip .4truecm
{\small \em $^b$Departamento de F\'{\i}sica, Centro de
	Investigaci\'on y de Estudios Avanzados del IPN}\\
{\small\em P.O. Box 14-740, CP. 07000, Ciudad de M\'exico, Mexico}\\

\vspace{0.3cm}

\vspace*{1cm}
\end{center}

\begin{abstract}

 \vskip .5truecm
Inspired by the formalism that relates the star-exponential with the quantum propagator for bosonic systems, in this work we introduce the analogous extension for the fermionic case. In particular, we analyse the problem of calculating the star-exponential (i.e., the symbol of the evolution operator) for Fermi systems within the deformation quantization program. Grassmann variables and coherent states are considered in order to obtain a closed-form expression for the fermionic star-exponential in terms of its associated propagator. As a primary application, a fermionic version of the Feynman-Kac formula is derived within this formalism, thus allowing a straightforward calculation of the ground state energy in phase space. Finally, the method is validated by successfully applying it to the simple harmonic and driven Fermi oscillators, for which the results developed here provide a powerful alternative computational tool for the study of fermionic systems.
\end{abstract}
\bigskip

\newpage

\section{Introduction}
\label{S-Intro}
Deformation quantization (DQ) is an approach to analyse both the classical and the quantum formalisms in the same dynamical space, namely, the phase space \cite{Bayen:1977ha,Bayen:1977hb}. Inspired by the development of phase-space quantum mechanics (PSQM), it is based on extending the algebra of observables into a more general structure that connects to classical phase-space quantities asymptotically. One of its most important elements is the existence of star-products that generalize the standard pointwise product of classical mechanics by introducing a noncommutative and associative algebraic structure that appropriately describes the inherent quantum mechanical framework. For decades, the only known example was the Moyal star-product for flat symplectic manifolds \cite{Bayen:1977ha,Bayen:1977hb}. The existence of star-products for general symplectic manifolds was proved in Refs. \cite{Wilde-Lecomte,Omori-1991,Fedosov:1994zz,Fedosov:1996fu}. Later a new star-product for Poisson manifolds was discovered in \cite{Kontsevich:1997vb}.

DQ is a versatile method of quantization which does not necessarily require the Hilbert space canonical structure, yielding nonetheless the correct spectrum of the quantum mechanical operators in an autonomous way (for some reviews see, \cite{Todorov:2012gy, Curtright:2014sli,Hirshfeld:2002yjb}). In fact, it yields the free and interacting spectra without requiring the traditional operator machinery in a proper Hilbert space. However, DQ confronts several technical issues whenever we invoke time evolution which, in both quantum mechanics and quantum field theory, are commonly described by path integrals via Feynman's propagators \cite{feynman2010quantum,Weinberg:1995mt,das2019field,Zinn-Justin-book-2004}. In DQ formalism the evolution is implemented by the {\it star-exponential} function (the symbols associated to the evolution operator). However, this function is still completely misunderstood due to convergence problems. In fact, the issue of convergence of general star-exponentials remains an open mathematical problem. Only in very few cases the star-exponential has been analytically obtained \cite{Bayen:1977ha,Bayen:1977hb,Dito:2005xt}, e.g., for the harmonic oscillator and the damped harmonic oscillator. The star-exponentials enters in some other contexts as it is exposed, for example, in Refs. \cite{cadavid-1991,Omori-2004,Omori:2012xv,Bieliavsky:2013eca,Yoshioka-2018,Yoshioka:2019tjf,Berra-Montiel:2025vhf}.

In Ref. \cite{Berra-Montiel:2024ubb} the connection of star-exponentials and path integrals was studied and established for bosonic systems. There, a mechanism that obtains the star-exponential function without relying on its explicit formal series calculation was devised: a closed-form integral relation equating the star-exponential with its quantum propagator. This is a direct and useful application that alleviates the problem of convergence by using an a priori known physical quantity (the quantum propagator) to calculate its associated symbol. Its potential applications are multiple, one of the most important being its connection to the Feynman-Kac formula \cite{Berra-Montiel:2025iuy}. The original formalism, however, was constrained to the bosonic case, thus limiting its applications to a very specific type of systems. Via the spin-statistics theorem, the existence and physical relevance of fermions is guaranteed, at least within relativistic quantum field theories in 3+1 dimensions. The classical dynamics and quantization of fermionic systems have been developed over the years \cite{Casalbuoni:1976tz,Casalbuoni:1975bj,Berezin:1976eg,Marnelius:1988pj,Lahiri:1990sv}. Moreover, the DQ of these systems has been also explored in Refs. \cite{Hirshfeld:2002ki,Hirshfeld:2004aj,Galaviz:2006ni,Odendahl:2007gzs,Lin:2022nns,Mrowczynski:2012ps}.

In the present article, we explore the aforementioned relation between the star-exponential and the Feynman propagator studied in \cite{Berra-Montiel:2024ubb}, as well as a direct application yielding the Feynman-Kac formula for determining the energy of the ground state for a fermionic system. We thus obtain a well-defined star-exponential together with the complementary Feynman-Kac formula in complete analogy to the bosonic systems discussed in \cite{Berra-Montiel:2025iuy}.

This work is organized as follows: In Section 2 we give some preliminary material associated with the Weyl-Wigner-Groenewold-Moyal formalism for fermionic quantum mechanics \cite{Galaviz:2006ni}. Section 3 is devoted to exploring the relation between the star-exponential and Feynman's propagator in quantum mechanics for fermions. In addition, some basic fermionic examples are provided for which one is able to explicitly construct the star-exponential function. In Section 4 the Feynman-Kac formula for fermionic systems and some examples are discussed. In Section 5, some final remarks and conclusions are given. Finally, an appendix is included in order to introduce the calculations for the ground state energies for the driven Fermi oscillator studied in Section 4.2.
\vskip 1truecm
\section{Brief Overview on Phase Space Fermionic Quantum Mechanics}
\label{Pre-LIM} 
In the present section we provide a very short overview on the tools we will employ in the subsequent sections. Our aim is not to give a detailed overview of deformation quantization of fermionic systems but just to introduce the notation and conventions we will follow throughout this article. A more precise and detailed introduction can be obtained from \cite{Galaviz:2006ni} and references therein, from where we mainly follow the notation and conventions.

First we introduce the phase space of fermionic variables which we define as $\Gamma^{2n}_{F}:= \{(\psi,\pi)\}$, where $\psi=(\psi_{1},\ldots,\psi_{n})$, $\pi=(\pi_{1},\ldots,\pi_{n})$, with $\psi_j$ and $\pi_k$ being complex Grassmann coordinates on $\Gamma^{2n}_{F}$ and $i,j,k,$ etc $\in \{1,\ldots,n\}$. The canonical momenta $\pi_{j}$ can be written in terms of $\psi$ in the form $\pi_{j}=\mathrm{i}\psi^{*}_{j},$ where $\psi^{*}$ denotes the complex conjugate of $\psi$. Canonical quantization rules establish
\begin{align}
[\widehat{\psi}_{j},\widehat{\pi}_{k}]_{+} &= \mathrm{i}\hbar\delta_{jk}, \label{anticom} \nonumber\\ 
[\widehat{\psi}_{j},\widehat{\psi}_{k}]_{+} &= 0 = [\widehat{\pi}_{j},\widehat{\pi}_{k}]_{+}, \nonumber\\
[\widehat{\psi}_{j},\widehat{\psi}^{*}_{k}]_{+} &= \hbar\delta_{jk},
\end{align}
where $[\cdot,\cdot ]_+$ is the anticommutator operation.

The operators are linear bounded operators acting on the Hilbert space ${\cal H} = L^2(\mathbb{R}^n) \otimes \mathbb{C}^2$. The proper creation and annihilation operators can be built as
$\widehat{b}_{j}:={\widehat{\psi}_{j}}/{\sqrt{\hbar}},$ and $\widehat{b}^{*}_{j}:={\widehat{\psi}^{*}_{j}}/{\sqrt{\hbar}}$. Thus the vacuum state is given by either the ket  $|0\rangle$ or by the bra  $\langle 0|$  by means of
$\widehat{b}_{j}|0\rangle=0,$ $\langle 0|\widehat{b}^{*}_{j}=0,$
for all $j$, where such states fulfil the normalization condition $\langle 0|0\rangle = 1$. The basis of all states can be constructed from excitations of the vacuum state $|0\rangle$ and it is given by
$|j,k,l,\ldots\rangle := \widehat{b}_j^*\widehat{b}_k^*\widehat{b}_l^*\cdots|0\rangle.$ State vectors $|j,k,l,\ldots\rangle$ fulfilling Pauli's exclusion principle should belong to the Fock space given by ${\cal F}({\cal H}) = \bigoplus_{k=0}^\infty {\cal A}({\cal H}^{\otimes k})$, where ${\cal A}$ is the antisymmetrization operator acting on the tensor products of ${\cal H}$. Moreover, states $|j,k,l,\ldots\rangle$ obey the relations $
\widehat{b}_l|j,k,\ldots\rangle = 0,$ $\widehat{b}_l^*|j,k,\ldots\rangle = |l,j,k,\ldots\rangle,$
and $\widehat{b}_l|l,j,k,\ldots\rangle = |j,k,\ldots\rangle,$ $\widehat{b}_l^*|l,j,k,\ldots\rangle = 0,$
where $l \notin \{j,k,\ldots\}$. For the dual basis one has $\langle j,k,l,\ldots| := \langle 0|\cdots\widehat{b}_l\widehat{b}_k\widehat{b}_j.$
Both states  $|j,k,l,\ldots\rangle$ and $\langle j,k,l,\ldots|$ can be used to form the inner product and verify that
\begin{equation}
\langle j_1,k_1,l_1,\ldots|j_2,k_2,l_2,\ldots\rangle = 
\begin{cases}
0 & \text{if } \{j_1,k_1,l_1,\ldots\} \neq \{j_2,k_2,l_2,\ldots\}\\
1 & \text{if } j_1 = j_2,k_1 = k_2,l_1 = l_2,\ldots
\end{cases}.
\end{equation}

\vskip .5truecm

Now, we look for the state $|\psi_1,\ldots,\psi_n\rangle \equiv |\psi\rangle$ that satisfies the following (eigenvalue) conditions
\begin{equation}
\widehat{\psi}_j|\psi\rangle = \psi_j|\psi\rangle,     \ \ \ \ \  \widehat{\psi}_j |0\rangle =0,
\end{equation}
for all $j$. Thus one can see that $|\psi\rangle$ has the following form
\begin{equation}
|\psi\rangle = \exp\left\{-\frac{\mathrm{i}}{\hbar}\sum_{j=1}^n \widehat{\pi}_j\psi_j\right\}|0\rangle.
\end{equation}
Then the eigenvalue equation is fulfilled $\widehat{\psi}_j|\psi\rangle = \psi_j|\psi\rangle,$  and  $\widehat{\psi}_j |0\rangle =0,$ for all $j$. Combining the previous results it is not difficult to get the crucial translation formula
\begin{equation}\label{translation}
\exp\left\{-\frac{\mathrm{i}}{\hbar}\sum_{j=1}^n\widehat{\pi}_j\xi_j\right\}|\psi\rangle = |\psi+\xi\rangle.
\end{equation}
In order to define the inner product $\langle \psi'|\psi \rangle$ let us find the corresponding bra, the dual vector to the ket. We seek a solution of the form $\langle\psi|\widehat{\psi}_{j}=\langle\psi|\psi_{j}$
for all $j$. After a lengthy but straightforward procedure one can deduce the bra state reads\footnote{The ordering in the product of the $\widehat{\psi}_{j}$'s can be arbitrarily chosen, and thus, we fix it to $\widehat{\psi}_{1}\cdots\widehat{\psi}_{n}$, for simplicity.} \cite{Galaviz:2006ni}
\begin{align}
\langle\psi|:=\langle 0|\widehat{\psi}_{1}\cdots\widehat{\psi}_{n} \exp\left\{-\frac{\mathrm{i}}{\hbar}\sum_{j=1}^{n}\psi_{j}\widehat{\pi}_{j}\right\} \nonumber \\
=\langle 0|\left(\prod_{j=1}^{n}\widehat{\psi}_{j}\right)\exp\left\{\frac{\mathrm{i}}{\hbar}\sum_{j=1}^{n}\widehat{\pi}_{j}\psi_{j}\right\}.
\end{align} 
From a straightforward computation we can see that $\langle\psi|\neq(|\psi\rangle)^{*}$. We can also see that the ket $|\psi\rangle$ commutes with all Grassmann numbers $\eta$, i.e., $\eta|\psi\rangle=|\psi\rangle\eta$. But for the bra defined for $\langle\psi|$ one arrives at
\begin{equation}
\eta\langle\psi|=(-1)^{e_{\eta}\cdot n}\langle\psi|\eta,
\end{equation}
where $e_{\eta}$ is the Grassmann parity of the number $\eta$. It is defined such as $e_{\eta}=1$ for odd Grassmann numbers and $e_{\eta}=0$ for even Grassmann numbers. The inner product $\langle\psi^{\prime}|\psi\rangle$ satisfies
$$
\langle\psi^{\prime}|\psi\rangle=\langle 0|\widehat{\psi}_{1}\cdots\widehat{\psi}_{n}\exp\left\{\frac{\mathrm{i}}{\hbar}\sum_{j=1}^{n}\widehat{\pi}_{j}\Bigl{(}\psi^{\prime}_{j}-\psi_{j}\Bigr{)}\right\}|0\rangle
$$
\begin{align}\label{delta psi}
=\prod_{j=1}^{n}\Bigl{(}\psi_{j}-\psi^{\prime}_{j}\Bigr{)}
=:\delta(\psi-\psi^{\prime}).
\end{align}
In the case of the $\pi$ basis one can find the following analogous results\footnote{For the case of the basis of momentum we have that the definition of the states $|\pi\rangle$ are given by $|\pi\rangle=\exp\left\{-\frac{\mathrm{i}}{\hbar}\sum_{j=1}^{n}\widehat{\psi}_{j}\pi_{j}\right\}\prod_{j=1}^{n}\widehat{\pi}_{j}|0\rangle,$ and the dual vector is written as: $\langle\pi|=\langle 0|\exp\left\{-\frac{\mathrm{i}}{\hbar}\sum_{j=1}^{n}\pi_{j}\widehat{\psi}_{j}\right\}=\langle 0|\exp\left\{\frac{\mathrm{i}}{\hbar}\sum_{j=1}^{n}\widehat{\psi}_{j}\pi_{j}\right\}.$ Self-evidently  $\langle\pi|\neq(|\pi\rangle)^{*}$. The inner product $\langle\pi^{\prime}|\pi\rangle$ yields:
$\langle\pi^{\prime}|\pi\rangle=\prod_{j=1}^{n}\left(\pi^{\prime}_{j}-\pi_{j}\right)
=\delta(\pi^{\prime}-\pi).$}. Mixing kets and bras of $\psi$ and $\pi$ one finds 
\begin{equation}
\langle\pi|\psi\rangle=\exp\left\{-\frac{\mathrm{i}}{\hbar}\sum_{j=1}^{n}\pi_{j}\psi_{j}\right\}, \ \ \ \ \
\langle\psi|\pi\rangle=(\mathrm{i}^n\hbar)^n \exp\left\{\frac{\mathrm{i}}{\hbar}\sum_{j=1}^n \pi_j \psi_j\right\}.
\end{equation}

From now on, we will adhere to the integral conventions from Weinberg's book \cite{Weinberg:1995mt}:
$\int \psi_j d\psi_j = - \int d\psi_j \psi_j = 1,$  $\int \pi_j d\pi_j = - \int d\pi_j \pi_j = 1$,
for all $j$, which yield the completeness condition
\begin{equation} \label{normalization}
\int |\psi\rangle \mathcal{D}\psi \langle \psi| = 1,
\end{equation}
where $\mathcal{D}\psi = d\psi_n \cdots d\psi_1$. Analogously a similar form to (\ref{normalization}) can be find for the $|\pi\rangle$ basis.

\subsection{Weyl-Wigner-Groenewold-Moyal Correspondence}
Let us start by considering Weyl's map for an arbitrary smooth function $f$ in phase space $\Gamma_F^{2n}$, let $f =  f(\pi,\psi) \in C^\infty(\Gamma_F^{2n})$ be an arbitrary classical observable:
\begin{equation}
    Q_{W}(f) = \int \widetilde{f}(\lambda,\mu) \exp \left\{ \mathrm{i} \sum_{j=1}^n (\widehat{\pi}_j \lambda_j + \widehat{\psi}_j \mu_j) \right\} \prod d\lambda d\mu,
\end{equation}
where $Q_W$ is an isomorphism map (called Weyl's transform) $Q_W: C^\infty(\Gamma_F^{2n}) \to {\cal B}({\cal H})$, where ${\cal B}({\cal H})$ is the set of quantum mechanical observables, which are linear bounded operators acting on a suitable Hilbert space ${\cal H}$. Moreover $\prod d\lambda d\mu := d\lambda_1 d\mu_1 \cdots d\lambda_n d\mu_n,$ while $\widetilde{f}(\lambda,\mu)$ stands for the Fourier transform of the function $f$:
\begin{equation}
\widetilde{f}(\lambda, \mu) := \int f(\pi, \psi) \exp \left\{ -\mathrm{i} \sum_{j=1}^n (\pi_j \lambda_j + \psi_j \mu_j) \right\} \prod d\pi d\psi,
\end{equation}
and $\prod d\pi d\psi := d\pi_1 d\psi_1 \cdots d\pi_n d\psi_n$. Using this explicit expansion of the transform and considering $f$ to be a smooth function on $\Gamma^{2n}_{F}$ we arrive at
\begin{equation}\label{f hat}
    Q_{W}(f) = \int f(\pi,\psi) \widehat{\Omega}(\pi,\psi) \prod d\pi d\psi,
\end{equation}
where the operator $\widehat{\Omega}(\pi,\psi)$ corresponds to the {\it Stratonovich-Weyl operator/quantizer}, given by
\begin{equation}
\widehat{\Omega}(\pi, \psi) = \int \exp \left\{ \mathrm{i} \sum_{j=1}^n [(\widehat{\pi}_j - \pi_j)\lambda_j + (\widehat{\psi}_j - \psi_j)\mu_j] \right\} \prod d\lambda d\mu.
\end{equation}
Using the translation formula (\ref{translation}) and the normalization equations (\ref{normalization}) we find, after some calculations, the following equivalent expression
\begin{align}
\widehat{\Omega}(\pi, \psi) &= 
 \mathrm{i}^n \int \exp \left\{ -\mathrm{i} \sum_{j=1}^n \pi_j \lambda_j \right\} \left| \psi - \frac{\hbar \lambda}{2} \right\rangle \left\langle \psi + \frac{\hbar \lambda}{2} \right| D\lambda.
\end{align}
In the $\pi-$basis it takes the form
\begin{align}
\widehat{\Omega}(\pi, \psi) &= (-1)^n \int \widehat{\Omega}(\pi, \psi) |\pi'\rangle \langle \pi'| \mathcal{D}\pi' \nonumber \\
&= (-i)^n \int \exp \left\{ -\mathrm{i} \sum_{j=1}^n \psi_j \mu_j \right\} \left| \pi - \frac{\hbar \mu}{2} \right\rangle \left\langle \pi + \frac{\hbar \mu}{2} \right| \mathcal{D}\mu.
\end{align}
Define the trace for any operator $\widehat{\cal O}$ as follows
\begin{eqnarray}
    \operatorname{tr}\{\widehat{\cal O}\} & = &  
    (\mathrm{i}\hbar)^{-n} \int \langle \psi | \widehat{\cal O} | \psi \rangle \mathcal{D}\psi  
     = (\mathrm{i}\hbar)^{-n} \int \langle \pi | \widehat{\cal O} | \pi \rangle \mathcal{D}\pi \,,  \nonumber\\
     \operatorname{tr}\{\widehat{\cal O}\widehat{\cal O}'\} 
&= &  
(-1)^{\widehat{e}_{\cal O} \widehat{e}_{{\cal O}'}} \operatorname{tr}\{\widehat{\cal O}'\widehat{\cal O}\} \,,
\label{trace AB}
\end{eqnarray}
where $\widehat{e}_{\cal O}$ and $\widehat{e}_{{\cal O}'}$ are the Grassmann parities of $\widehat{\cal O}$ and $\widehat{{\cal O}'}$, respectively. With
the previous definition of trace, the Stratonovich-Weyl operator admits a well-defined trace and satisfies the following properties:
\begin{eqnarray}
\operatorname{tr}\{\widehat{\Omega}(\pi', \psi') \widehat{\Omega}(\pi'', \psi'')\} 
&=& 
(\psi_1' - \psi_1'')(\pi_1' - \pi_1'') \cdots (\psi_n' - \psi_n'')(\pi_n' - \pi_n'')  \nonumber \\
&=& \delta(\psi' - \psi'', \pi' - \pi'') \,,\\
    \operatorname{tr}\{\widehat{\Omega}(\pi,\psi)\} 
    & = &  1 \,.
\end{eqnarray}

Consider now Weyl's inverse quantization map
\begin{equation}
    Q_{W}^{-1}(\widehat{f}) = \operatorname{tr}\{\widehat{f}\widehat{\Omega}(\pi,\psi)\} \,.
\end{equation}
This relation defines a symbol $f$ from its quantization $\widehat{f}$. We will restrict our analysis to the case of integrable functions for simplicity. Employing (\ref{normalization}) and (\ref{trace AB}) one arrives at the following relation
\begin{align}
f(\pi, \psi) = \operatorname{tr}\{\widehat{f} \widehat{\Omega}(\pi, \psi)\} \nonumber\\ 
= \operatorname{tr}\{\widehat{\Omega}(\pi, \psi) \widehat{f}\}.
\end{align}
Using Weyl's quantization map and its inverse we can calculate the Wigner function corresponding to the density operator acting on the Hilbert space
$$
\rho_{W}(\pi, \psi) = \operatorname{tr}\{\widehat{\rho} \widehat{\Omega}(\pi, \psi)\}
    = (\mathrm{i}\hbar)^{-n} \int \langle \psi' | \widehat{\rho} \widehat{\Omega}(\pi, \psi) | \psi' \rangle \mathcal{D}\psi'
$$ 
\begin{equation}
    = \hbar^{-n} \int \exp \left\{ -\mathrm{i} \sum_{j=1}^{n} \pi_j \lambda_j \right\} \left\langle \psi + \frac{\hbar \lambda}{2} \middle| \widehat{\rho} \middle| \psi - \frac{\hbar \lambda}{2} \right\rangle \mathcal{D}\lambda.
\end{equation}
The Wigner function is a \textit{quasi probability} distribution, which can be used to determine the expectation values of operators by integrating their associated symbols over the Grassmann phase space
\begin{equation}
 \langle \widehat{\cal O} \rangle = \int \rho(\pi,\psi) {\cal O}(\pi,\psi) \prod d\pi d\psi,
\end{equation}
where ${\cal O} = Q_{W}^{-1}(\widehat{\cal O})$ is obtained using Weyl's inverse quantization map. Now let us overview star-products for the case of fermions.

\subsection{Star-product}
Using the bijectivity of Weyl's map and the closure of Hilbert-Schmidt operators we find that there is a unique function, $f \star g$, on $\Gamma^{2n}_{F}$ such that
\begin{equation}
    Q_{W}(f)Q_{W}(g) = Q_{W}(f \star g).
\end{equation}
This function is called the {\it Moyal star-product}, and it is explicitly given by
\begin{align}
 (f \star g)(\pi,\psi) &= Q_{W}^{-1}(Q_{W}(f)Q_{W}(g))  \nonumber\\
 &= \operatorname{tr} \left\{ \widehat{f}\widehat{g}\widehat{\Omega}(\pi,\psi) \right\}.
\end{align}
The details of the deduction of the following identity are provided in \cite{Galaviz:2006ni}. The trace of the product of three Stratonovich-Weyl operators is written as
\begin{align}\label{triple s-w}
\operatorname{tr}\{\widehat{\Omega}(\pi', \psi') \widehat{\Omega}(\pi'', \psi'') \widehat{\Omega}(\pi, \psi)\} 
&= \left( \frac{\mathrm{i}\hbar}{2} \right)^{2n} \exp \left\{ -\frac{2\mathrm{i}}{\hbar} (\pi'(\psi'' - \psi) + \pi''(\psi - \psi') + \pi(\psi' - \psi'')) \right\}.
\end{align}
Combining this last identity with (\ref{f hat}) inside the trace we find
\begin{align}
(f \star g)(\pi, \psi) &= \left( \frac{\mathrm{i}\hbar}{2} \right)^{2n} \int f(\pi', \psi') g(\pi'', \psi'') \nonumber\\
&\quad \times \exp \left\{ -\frac{2\mathrm{i}}{\hbar} [\pi'(\psi'' - \psi) + \pi''(\psi - \psi') + \pi(\psi' - \psi'')] \right\} \mathcal{D}\pi' \mathcal{D}\psi' \mathcal{D}\pi'' \mathcal{D}\psi''.
\end{align}
By introducing the following variables: $\Psi'' = \psi' - \psi$, $\Pi'' = \pi' - \pi$, $\Psi''' = \psi'' - \psi$, $\Pi''' = \pi'' - \pi$, the previous form for the Moyal product changes into
\begin{align}
(f \star g)(\pi, \psi) &= \left( \frac{\mathrm{i}\hbar}{2} \right)^{2n} \int f(\Pi' + \pi, \Psi'' + \psi) g(\Pi'' + \pi, \Psi'' + \psi) \nonumber \\
&\quad \times \exp \left\{ -\frac{2\mathrm{i}}{\hbar} [\Pi'' \Psi'' - \Pi'' \Psi'] \right\} \mathcal{D}\Pi' \mathcal{D}\Psi'' \mathcal{D}\Pi'' \mathcal{D}\Psi'',
\end{align}
which stands for the integral representation of the star-product.
\par The subsequent expansion of the functions $f$ and $g$ into a Taylor series \cite{Galaviz:2006ni} gives rise to a closed form for the Moyal product
\begin{equation}
(f \star g)(\pi, \psi) = f(\pi, \psi) \exp \left\{ \frac{\mathrm{i}\hbar}{2} \overset{\leftrightarrow}{\mathcal{P}}_F \right\} g(\pi, \psi) \,,
\end{equation}
where $\overset{\leftrightarrow}{\mathcal{P}}_F$ is the Fermi-Poisson tensor, which plays the role of a Poisson structure for Grassmann variables, and it may be explicitly written as
\begin{equation}\label{eq:PF}
\overset{\leftrightarrow}{\mathcal{P}}_F = \frac{\overleftarrow{\partial}}{\partial \pi} \frac{\overrightarrow{\partial}}{\partial \psi} + \frac{\overleftarrow{\partial}}{\partial \psi} \frac{\overrightarrow{\partial}}{\partial \pi},
\end{equation}
with \(\overrightarrow{\partial}\) and \(\overleftarrow{\partial}\) denoting the right and left derivatives, respectively. This structure naturally generalizes the standard Moyal product to the fermionic domain. For a more complete description of the Poisson structure for systems involving bosonic and fermionic variables see \cite{Casalbuoni:1976tz,Casalbuoni:1975bj,Berezin:1976eg,Lahiri:1990sv}. The Poisson bracket for fermionic systems corresponds to the odd part of a symplectic structure on a symplectic superspace (or supermanifold) (for more details see for instance, \cite{DeWitt:2012mdz}).

\par With these basic tools established, let us explore in the following sections their relation with propagators within the path integral approach and with the Feynman-Kac formula in the fermionic context.
\vskip 1truecm
\section{Star-exponential from Propagators for Fermions}
In reference \cite{Berra-Montiel:2024ubb} the relation between the star-exponential and path integrals for the bosonic case was established. In the present section we will carry out the mentioned correspondence for fermionic systems. It is known that in the fermionic case \cite{Galaviz:2006ni} the following equality holds: $\bra{\pi} =(\ket{\psi})^{*}$. The form of the fermionic propagator most convenient for our purposes in this article is the one discussed in \cite{Engl:2014rwj}
\begin{equation}
K(\psi_{f},t;\psi_{0},0) = \bra{\pi_{f}}e^{-\frac{i}{\hbar}t\widehat{H}}\ket{\psi_{0}}.
\end{equation}
Then, the Wigner function can be written as
\begin{equation}
    \rho_{f,o}(\psi,\pi) = Q_{W}^{-1}(\ket{\psi_{0}}\bra{\pi_{f}}).
\end{equation}
Using results from \cite{Galaviz:2006ni} one may find that
\begin{align}
   \label{density}
    \rho_{f,o}(\pi,\psi)  &=  \hbar^{-n} \int \exp\bigg\{-i\sum_{j=1}^{n}\pi_{j}\lambda_{j}\bigg\}\bra{\psi + \frac{\hbar\lambda}{2}}\widehat{\rho}\ket{\psi-\frac{\hbar\lambda}{2}} D\lambda  \nonumber \\
    &= \hbar^{-n}\int \exp\bigg\{-i\sum_{j=1}^{n}\pi_{j}\lambda_{j}\bigg\}\bra{\psi + \frac{\hbar\lambda}{2}}\ket{\psi_{0}}\bra{\pi_{f}}\ket{\psi - \frac{\hbar\lambda}{2}} D\lambda \nonumber \\
    &=\hbar^{-n} \int \exp\bigg\{-i\sum_{j}\pi_{j}\lambda_{j}\bigg\} \delta\bigg(\psi_{0}-\bigg(\psi+\frac{\hbar\lambda}{2}\bigg)\bigg)\\ \nonumber
   & \times \exp\bigg\{-\frac{i}{\hbar}\sum_{j}\pi_{j,f}\bigg(\psi_{j}-\frac{\hbar\lambda_{j}}{2}\bigg)\bigg\} D\lambda,
\end{align}
where $\widehat{\rho}= |\psi_0\rangle \langle \pi_f|$, and where the subscript $f$ in $\pi_{j,f}$ denotes the final momentum. Now using the formula for Grassmann variables $\delta(f(\theta)) = \sum_{i}\delta(\theta-\theta_{i})f^{'}(\theta_{i})$, where $\theta_{i}$ stands for the zeros of $f$ in the argument of the delta symbol, and using new variables $\chi_j := \psi_j - \psi_0$ and $\eta_{j} := \pi_{j} - \pi_{j,f}$, the integration in (\ref{density}) can be carried out and yields 
\begin{align}
    \rho_{f,0} = 2^{-n}\exp\left\{\frac{2i}{\hbar}\eta \cdot \chi\right\}\exp\left\{-\frac{i}{\hbar}\pi_{j,f}\cdot \psi_{j}^{0}\right\}.
\end{align}
Now, for fermionic quantum mechanics the propagator is given by
$$
K(\pi_{f}, t ; \psi_{0},0) = \bra{\pi_{f}} \exp\left\{-\frac{i}{\hbar}t\widehat{H}\right\}\ket{\psi_{0}} = \int \rho_{f,0}(\psi,\pi) {\rm Exp}_{\star}\left\{-\frac{it}{\hbar}H(\psi,\pi)\right\}D\psi D\pi
$$
\begin{equation}= 2^{-n}\exp\left\{-\frac{i}{\hbar}\pi_{j,f}\cdot \psi_{j}^{0}\right\} \int \exp\left \{\frac{2i}{\hbar}\eta \cdot \chi\right\}{\rm Exp}_{\star}\left \{-\frac{it}{\hbar}H(\chi + \psi^{0},\eta + \pi_{f})\right\} D\chi D\eta,
\label{propagator}
\end{equation}
where ${\rm Exp}_{\star}$ is the star-exponential, which corresponds to the symbol (the inverse Weyl-Wigner map) of the evolution operator. In order to solve for the star-exponential inside the integral we consider the following change of variables
\begin{align}
    \Psi = \frac{\psi_{f}+\psi_{0}}{2} , \ \ \ \   \Psi^{'} = \frac{\psi_{f}-\psi_{0}}{2}, \ \ \ \ \Pi = \frac{\pi_{f}+\pi_{0}}{2}, \ \ \ \ \Pi^{'} = \frac{\pi_{f}-\pi_{0}}{2}.
\end{align}
Consider the following quantity
\begin{align}
\widetilde{F}(\Psi,\Pi)= C\exp\left\{\frac{i}{\hbar}\Pi\cdot\Psi\right\}\int \exp\left\{-\frac{2i}{\hbar}\Pi^{'}\cdot\Psi^{'}\right\}K(\Pi +\Pi^{'},t;\Psi-\Psi^{'}) D\Psi^{'}D\Pi^{'},
\label{dos}
\end{align}
where $C$ is a constant that ensures the normalization of the result. Now, substituting (\ref{propagator}) into (\ref{dos}) and using new variables $ \chi = \chi^{'}+\Psi{'}$ and $ \eta = \eta^{'}-\Pi^{'}$ we are able to express (\ref{dos}) as

\begin{align}
    \widetilde{F}(\Psi,\Pi) = 2^{-n}C\int \exp\left\{\frac{2i}{\hbar}\eta^{'}\cdot\chi^{'}\right\} {\rm Exp}_{\star}\left\{-\frac{it}{\hbar}H(\chi^{'} + \Psi,\eta^{'} + \Pi)\right\} \widetilde{G} D\chi^{'} D\eta^{'},
\end{align}
where
\begin{align}
    \widetilde{G} := \int \exp\bigg\{\frac{i}{\hbar}\bigg[-3\Pi^{'}\cdot\Psi^{'}-\Psi^{'}\cdot(2\eta^{'}+\Pi)-\Pi^{'}\cdot(\Psi + 2\chi^{'})\bigg]\bigg\} D\Psi^{'}D\Pi^{'}.
\end{align}
Now using the Gaussian integral in the form
\begin{align}
    J = \int \exp(\mathbf{v}^T M \mathbf{u} + \mathbf{u}^T \mathbf{a} + \mathbf{v}^T \mathbf{b}) D\mathbf{u} \, D\mathbf{v}  \nonumber \\
    = \det(M) \, \exp(-\mathbf{a}^T M^{-1} \mathbf{b})
\end{align}
and taking $M = \frac{i}{\hbar}(-3) \mathrm{I}$, $a = -\frac{i}{\hbar}(2\eta^{'}+\Pi)$, $b= -\frac{i}{\hbar}(\Psi + 2 \chi^{'})$, $M^{-1} = \frac{\hbar i}{3} \mathrm{I}$, with ${\rm det}(M) = \left(-\frac{3i}{\hbar}\right)^{n},$ $-a^{T}M^{-1}b = \frac{i}{3\hbar}(2\eta^{'}+\Pi)\cdot(\Psi+2\chi^{'}),$ where I is the unity operator, we get
\begin{align}
    \widetilde{G} = \bigg(-\frac{3i}{\hbar}\bigg)^{n}\exp\left\{\frac{i}{3\hbar}(2\eta^{'}+\Pi)\cdot(\Psi+2\chi^{'})\right\},
\end{align}
and finally we obtain
\begin{align}
    \widetilde{F}(\Psi,\Pi)= 2^{-n} C (-\frac{3i}{\hbar})^{n} \int e^{{S}} {\rm Exp}_{\star}\left\{-\frac{it}{\hbar}H(\chi^{'}+\Psi,\eta^{'}+\Pi)\right\} D\chi^{'}D\eta^{'},
\end{align}
where
\begin{align}
    {S} =\frac{i}{\hbar}\left(\frac{10}{3}\eta{'}\cdot\chi{'}+\frac{2}{3}\eta{'} \cdot \Psi + \frac{1}{3}\Pi\cdot\Psi + \frac{2}{3}\Pi\cdot\chi{'}\right).
\end{align}
Expanding the exponential $e^S$ up to second order given its Grassmann nature and using the top-form of Grassmann integrals one finds
\begin{align}
    \widetilde{F}(\Psi,\Pi) = &C\bigg(-\frac{3i}{2\hbar}\bigg)^{n} \int \left[1 +\frac{i}{\hbar}\bigg(\frac{10}{3}\eta{'}\cdot\chi{'}+\frac{2}{3}\eta{'} \cdot \Psi + \frac{1}{3}\Pi\cdot\Psi + \frac{2}{3}\Pi\cdot\chi{'}\bigg) \right] \nonumber \\
    &\times {\rm Exp}_{\star}\left(-\frac{it}{\hbar}H(\chi'+\Psi, \eta^{'}+\Pi)\right)  D\chi'D\eta' \nonumber \\
    &= {\rm Exp}_{\star}\left(-\frac{it}{\hbar}H(\Psi, \Pi)\right),
\end{align}
where
\begin{equation}
C = \left( \frac{2^{n-1} \hbar^{n+1}}{5 \cdot 3^{n-1}} \right) i^{3-3n}.
\end{equation}
Thus the star-exponential can be recast in terms of the fermionic propagator in the form
\begin{align}
   {\rm Exp}_{\star}\left\{-\frac{it}{\hbar}H(\Psi,\Pi)\right\} =i^{3-3n}  \left( \frac{2^{n-1} \hbar^{n+1}}{5 \cdot 3^{n-1}} \right) \exp\left\{\frac{i}{\hbar}\Pi\cdot\Psi\right\}  \nonumber \\
   \times\int  \exp\left\{-\frac{2i}{\hbar}\Pi^{'}\cdot\Psi^{'}\right\}K(\Pi +\Pi^{'},t;\Psi-\Psi^{'}) D\Psi^{'}D\Pi^{'}.
\label{exponential-star-general}
\end{align}
This expression for the star-exponential thus generalizes the analogous identity developed in~\cite{Berra-Montiel:2024ubb}. As discussed there, the star-exponential obtained through the propagator may offer a potential advantage, as this expression avoids convergence issues emerging in the common definition of the star-product. Let us now see how it can be applied to different simple fermionic systems.

\subsection{Fermi Oscillator}
The discussion of this subsection is adapted from \cite{das2019field} (see also, \cite{Zinn-Justin-book-2004}). Let $\psi$ and $\bar{\psi}$ be two independent time-dependent Grassmann variables. The oscillator behavior can be represented by the following Lagrangian
\begin{align}\label{L1}
    L_{1} = \frac{i}{2}(\bar{\psi}\dot{\psi}-\dot{\bar{\psi}}\psi) - \frac{\omega}{2}[\bar{\psi}, \psi],
\end{align}
where $\omega$ denotes the oscillation frequency and $[\cdot,\cdot]$ denotes the commutator. A straightforward calculation shows that the following Lagrangian is equivalent (up to a total derivative) to
\begin{align}
    L_{2} = i\bar{\psi}\dot{\psi} - \frac{\omega}{2}[\bar{\psi}, \psi].
\end{align}
For simplicity, let us consider $L_{1}$. We begin by calculating the canonical conjugate momenta
\begin{align}
    \Pi_{\psi} = \frac{\partial L_{1}}{\partial\dot{\psi}} = -\frac{i}{2}\bar{\psi}, \ \ \ \ \  \Pi_{\bar{\psi}} = \frac{\partial L_{1}}{\partial \dot{\bar{\psi}}} = -\frac{i}{2}\psi.
\end{align}
We calculate its Hamiltonian via the Legendre transform
\begin{align}
    H = \frac{\omega}{2} [\bar{\psi},\psi].
\end{align}

Following our convention these independent Grassmann variables are complex. This allows us to identify the quantization maps
\begin{align}
    \psi \quad\rightarrow\quad\widehat{\psi} \quad ,\quad 
    \bar{\psi} \quad\rightarrow\quad i \widehat{\pi} \,,
\end{align}
which implies that the quantum Hamiltonian operator takes the form
\begin{align}
    \widehat{H} =\frac{i\omega}{2}[\widehat{\pi}, \widehat{\psi}].
\end{align}
Using the anticommutation relations (\ref{anticom}), we obtain
\begin{align}
    \widehat{H} &= \frac{i\omega}{2}[\widehat{\pi}, \widehat{\psi}] 
    = \frac{i\omega}{2}(\widehat{\pi}\widehat{\psi} - \widehat{\psi}\widehat{\pi}) \nonumber\\
    &=\frac{i\omega}{2}(\widehat{\pi}\widehat{\psi} + (\widehat{\pi}\widehat{\psi} -i\hbar)) 
    = \omega\left(i\widehat{\pi}\widehat{\psi}+\frac{\hbar}{2}\right).
\end{align}
Let us now calculate the evolution operator
\begin{align}
    \widehat{U}(t,t_{0}) &= \exp\{-i\widehat{H}(t-t_{0})\} \nonumber \\
    &= \exp\left\{\omega\widehat{\pi}\widehat{\psi}(t-t_{0})-\frac{i\omega\hbar}{2}(t-t_{0})\right\} \nonumber\\
    &=\exp\left\{\omega\widehat{\psi}\widehat{\pi}(t-t_{0})\right\}\exp\left\{-\frac{i\omega\hbar}{2}(t-t_{0})\right\} \nonumber\\
    &= \exp\left\{\omega\widehat{\psi}\widehat{\pi}(t-t_{0})\right\} \times [\text{phase}].
\end{align}
As it is well known, the Hamiltonian of a physical system is invariant under a global phase shift; therefore, it is possible to adjust this phase term. We will proceed in this manner for the propagator in the following section.

Following the step-by-step derivation detailed in \cite{Kafuri:2026icq}, or alternatively by working in a two-dimensional basis analogous to the fermionic oscillator example in \cite{Galaviz:2006ni}, one finds that the propagator $K$ satisfies the following formula
\begin{align}
    K(\pi_{f},t;\psi_{0},0) &= \exp \left\{\frac{i\omega t}{2}\right\}\exp\{\pi_{f}e^{-i\omega t}\psi_{0}\}  \nonumber \\
    &= \exp\left\{\frac{i\omega t}{2}\right\}\exp\{(\Pi +\Pi')e^{-i\omega t}(\Psi-\Psi')\}  \nonumber \\
    &= \exp\left\{\frac{i\omega t}{2}\right\} + [\Pi\Psi-\Pi\Psi'+\Pi'\Psi-\Pi'\Psi']\exp\left\{-\frac{i\omega t}{2}\right\},
\end{align}
where we have used the property $e^A = 1+A$ for nilpotent Grassmann bilinears. Similarly, the exponential kernel can be expanded up to first order terms as
\begin{align}
    e^{-i\frac{2}{\hbar}\Pi'\cdot\Psi'} = 1- \frac{2i}{\hbar}\Pi'\Psi'.
\end{align}
Multiplying the kernel by the propagator, we obtain
\begin{align}
    e^{-i\frac{2}{\hbar}\Pi'\Psi'}K &= \exp\left\{\frac{i\omega t}{2}\right\} + [\Pi\Psi-\Pi\Psi'+\Pi'\Psi-\Pi'\Psi']\exp\left\{-\frac{i\omega t}{2}\right\} \nonumber \\
    &\quad - \frac{2i}{\hbar}\Pi'\Psi'\exp\left\{\frac{i\omega t}{2}\right\} - \frac{2i}{\hbar}\Pi'\Psi'\Pi\Psi\exp\left\{-\frac{i\omega t}{2}\right\}.
\end{align}
We now perform the integration over the auxiliary variables $\Psi'$ and $\Pi'$. Using the standard Grassmann integration rules ($\int d\theta = 0, \int \theta d\theta = 1$), only terms containing the product $\Pi'\Psi'$ survive. This yields
\begin{align}
    \int \exp\left\{-\frac{2i}{\hbar}\Pi'\Psi'\right\}K D\Psi'D\Pi' = -\exp\left\{-\frac{i\omega t}{2}\right\} - \frac{2i}{\hbar}\Pi\Psi\exp\left\{-\frac{i\omega t}{2}\right\} - \frac{2i}{\hbar}\exp\left\{\frac{i\omega t}{2}\right\}.
\end{align}
Finally, substituting the normalization constant $C$ (which for $n=1$ corresponds to $-\hbar^2/5$ in our convention), we arrive at the star-exponential
$$
    {\rm Exp}_{\star}\left\{-\frac{it}{\hbar}H(\Psi,\Pi)\right\} = C\exp\left\{\frac{i}{\hbar}\Pi\Psi\right\} \int \exp\left\{-\frac{2i}{\hbar}\Pi'\Psi'\right\}K D\Psi'D\Pi'
$$    
\begin{equation}
   = - \left(\frac{\hbar^{2}}{5}\right) \Bigg[ \exp\left\{\frac{i}{\hbar}\Pi\Psi\right\}\left(\exp\left\{-\frac{i\omega t}{2}\right\} + \frac{2i}{\hbar}\exp\left\{\frac{i\omega t}{2}\right\}\right) + \frac{2i}{\hbar}\Pi\Psi\exp\left\{-\frac{i\omega t}{2}\right\} \Bigg].
\label{expstar}
\end{equation}
This exact expression highlights the advantage of the integral approach, as it circumvents the convergence issues typically encountered when attempting to compute the star-exponential directly via the formal power series of star-products. Moreover the star-exponential will be a very useful later to find the ground state energy of a system of fermions through the corresponding Feynman-Kac formula.  

\subsection{Driven Fermi Oscillator}
Consider a driven fermionic model defined by the Hamiltonian $\widehat{H} = \omega \widehat{\pi}\widehat{\psi} + \alpha \widehat{\psi}$, where $\omega$ is the frequency and $\alpha$ is a Grassmann-valued coupling parameter. Since this expression is not Hermitian, we must include the complex conjugate of the linear term to ensure a physical observable. The resulting Hamiltonian is
\begin{align}
    \widehat{H}=\omega \widehat{\pi}\widehat{\psi} + \alpha\widehat{\psi} + \alpha^{*}\widehat{\pi}.
\end{align}
In the occupation number basis $\{\ket{0}, \ket{1}\}$, the action of the operators is defined by:
\begin{align}
    \widehat{\psi}\ket{1} = \ket{0}, \quad \widehat{\psi}\ket{0} = 0, \nonumber \\
    \widehat{\pi}\ket{1} = 0, \quad \widehat{\pi}\ket{0} = \ket{1},
\end{align}
and the number operator $\widehat{N} = \widehat{\pi}\widehat{\psi}$ satisfies $\widehat{N}\ket{n} = n\ket{n}$. The matrix representations in this basis are given by
\begin{align}
    (\widehat{\psi}) = \begin{pmatrix}
    0 & 1 \\
    0 & 0
\end{pmatrix}, \quad 
    (\widehat{\pi}) = \begin{pmatrix}
    0 & 0 \\
    1 & 0
\end{pmatrix}, \quad 
    (\widehat{N}) = \begin{pmatrix}
    0 & 0 \\
    0 & 1
\end{pmatrix}.
\end{align}
This implies that the Hamiltonian takes the form
\begin{align}
    (\widehat{H}) &= \omega \begin{pmatrix}
        0 & 0 \\
        0 & 1
    \end{pmatrix} +\alpha \begin{pmatrix}
        0 & 1 \\
        0 & 0
    \end{pmatrix} + \alpha^{*} \begin{pmatrix}
        0 & 0 \\
        1 & 0
    \end{pmatrix} \nonumber\\
    &= \begin{pmatrix}
        0 & \alpha \\
        \alpha^{*} & \omega
    \end{pmatrix}.
\end{align}
To determine the ground state energy, we diagonalize the Hamiltonian. The characteristic equation arises from the condition $\det(\widehat{H} - \lambda \mathbb{I}) = 0$:
\begin{align}
    \det \begin{bmatrix}
        - \lambda & \alpha \\
        \alpha^{*} & \omega-\lambda
    \end{bmatrix} &= -\lambda(\omega-\lambda) - |\alpha|^{2} \nonumber\\
    &= \lambda^{2} - \lambda \omega - |\alpha|^{2} = 0.
\end{align}
Solving for the eigenvalues $\lambda$, we obtain
\begin{align}
    \lambda_{\pm} = \frac{\omega \pm \sqrt{\omega^{2}+ 4|\alpha|^{2}}}{2}.
\end{align}
The non-diagonal terms in the Hamiltonian mix states with different particle numbers, requiring a transformation to a new basis. The eigenvalues obtained correspond to the energies of the resulting quasi-particle excitations.

\subsubsection*{Heisenberg's Equations of Motion}
Let $\widehat{A}$ be an operator. Its time evolution is governed by Heisenberg's equation (setting $\hbar = 1$ for algebraic simplicity):
\begin{align}
    i \frac{d\widehat{A}}{dt} = [\widehat{A},\widehat{H}].
\end{align}
Recall the identity $[A, BC] = [A,B]C - B[A,C]$. Since $\alpha$ and $\alpha^{*}$ are Grassmann variables, they anticommute with the fermionic operators. For $\widehat{\psi}$, we have
\begin{align}
    [\widehat{\psi}, \widehat{H}] &= [\widehat{\psi}, \omega\widehat{\pi}\widehat{\psi} + \alpha \widehat{\psi} + \alpha^{*}\widehat{\pi}] \nonumber\\
    &= \omega[\widehat{\psi},\widehat{\pi}]\widehat{\psi} - \omega\widehat{\pi}[\widehat{\psi},\widehat{\psi}] + [\widehat{\psi},\alpha]\widehat{\psi} - \alpha[\widehat{\psi},\widehat{\psi}] + \{\widehat{\psi},\alpha^{*}\}\widehat{\pi} - \alpha^{*}[\widehat{\psi},\widehat{\pi}].
\end{align}
Using the canonical anticommutation relations $\{\widehat{\psi}, \widehat{\pi}\} = i$ and the nilpotency $\{\widehat{\psi}, \widehat{\psi}\} = 0$, this simplifies to
\begin{align}
    [\widehat{\psi}, \widehat{H}] = i(\omega\widehat{\psi} - \alpha^{*}).
\end{align}
Analogously, for $\widehat{\pi}$ one finds
\begin{align}
    [\widehat{\pi}, \widehat{H}] = i(-\omega\widehat{\pi} - \alpha).
\end{align}
This yields the following system of coupled linear differential equations
\begin{align}
    \dot{\widehat{\psi}} = -i(\omega\widehat{\psi} - \alpha^{*}), \quad \quad \dot{\widehat{\pi}} = -i(-\omega\widehat{\pi} - \alpha).
\end{align}
These are first-order linear inhomogeneous differential equations. Using the integrating factor method, the general solution for $\widehat{\psi}(t)$ with initial condition at $t=0$ is written as
\begin{align}
    \widehat{\psi}(t) = e^{-i\omega t}\widehat{\psi}(0) + \frac{\alpha^{*}}{\omega}(1-e^{-i\omega t}).
\end{align}
Similarly, for $\widehat{\pi}(t)$, applying the transformation $\omega \to -\omega$ and $\alpha^* \to \alpha$:
\begin{align}
    \widehat{\pi}(t) = e^{i\omega t}\widehat{\pi}(0) + \frac{\alpha}{\omega}(e^{i\omega t} -1).
\end{align}

\subsubsection*{Decomposition of the Propagator}
As established for the bosonic case \cite{Berra-Montiel:2024ubb}, the propagator takes the form $K = \mathcal{N}e^{iS_{cl}}$. Because the equations of motion are linear, the total classical action $S_{cl}$ can be split into a homogeneous part (dependent only on boundary conditions) and an inhomogeneous part (dependent on the sources). We decompose the calculation of the propagator into four distinct terms corresponding to the different sectors of the action:
\begin{enumerate}
    \item \textit{Homogeneous Propagator (Free Evolution)}\\
    In the absence of sources $(\alpha =\alpha^{*}=0)$, we recover the standard harmonic oscillator result. Note that we temporarily omit the global phase shift (vacuum energy) which will be restored at the end
    \begin{align}
        {\cal A}_{1} = \bar{\psi}_{f}e^{-i\omega t}\psi_{i}.
    \end{align}

    \item \textit{Source-Initial State Coupling (Linear Term 1)}\\
    This term represents the interaction between the final boundary condition $\bar{\psi}_f$ and the evolution of the initial state driven by the source. It is obtained by projecting the inhomogeneous part of $\psi(t)$ onto $\bar{\psi}_f$
    \begin{align}
        {\cal A}_{2} &= \bar{\psi}_{f} \cdot (\psi(t)_{\text{inhom}}) \nonumber \\
        &= \bar{\psi}_{f} \left[ \frac{\alpha^{*}}{\omega}(1-e^{-i\omega t}) \right] \nonumber \\
        &= - \frac{\alpha^{*}}{\omega}(1-e^{-i\omega t})\bar{\psi}_{f}.
    \end{align}

    \item \textit{Source-Final State Coupling (Linear Term 2)}\\
    This term arises from the interaction of the momentum source with the initial state $\psi_i$. It involves integrating the source term for $\widehat{\pi}$ against the time-reversed evolution kernel $D(t'-t) = e^{i\omega(t'-t)}$
    \begin{align}
        {\cal A}_{3} &= -i\int_{0}^{t}\alpha D(t'-t) dt' \nonumber\\
        &= -i\alpha e^{-i\omega t}\int_{0}^{t}e^{i\omega t'}dt' \nonumber\\
        &= - \frac{\alpha}{\omega}(1-e^{-i\omega t})\psi_{i}.
    \end{align}

    \item \textit{Source Self-Interaction (Constant Term)}\\
    Finally, we must calculate the term independent of the boundary fields $\psi_i$ and $\bar{\psi}_f$. This corresponds to the classical action of the sources interacting with themselves, often denoted as the ``vacuum-to-vacuum" amplitude in the presence of sources. Following the functional methods (see for instance,  Das \cite{das2019field}), the phase is given by
    \begin{align}
        \Phi(t) &= -\frac{i}{2}\int_{0}^{t}dt'\int_{0}^{t'}dt'' \left[ \bar{\eta}(t') G_F(t'-t'') \eta(t'') + \text{h.c.} \right],
    \end{align}
    where the effective sources are $\eta(t) = -i \alpha^{*}$ and $\bar{\eta}(t) = -i \alpha$, and where $G_F$ is the Feynman Green's function. Evaluating the double integral for constant sources yields
    \begin{align}
        \Phi(t) &= |\alpha|^{2}\int_{0}^{t}dt' e^{i\omega t'} \int_{0}^{t'}dt'' e^{-i\omega t''} + \text{const} \nonumber\\
        &= \frac{i|\alpha|^{2}}{\omega} \left[ t - \frac{1}{i\omega}(e^{i\omega t}-1) \right].
    \end{align}
    In order to obtain a global factor consistent with the fermionic trace, we define the function
    \begin{align}
        {\cal A}_{4} = f(\alpha, t, \omega) = \frac{|\alpha|^{2}}{2\omega}(-i\omega t + e^{-i\omega t} -1).
    \end{align}
\end{enumerate}

Collecting all four terms, the full propagator for the driven fermionic oscillator is
\begin{align}
    K(\pi_{f},t;\psi_{0}, 0) = \exp\left\{ \pi_{f}e^{-i\omega t}\psi_{0} - \frac{1}{\omega}(1-e^{-i\omega t})(\alpha^{*}\pi_{f} +\alpha\psi_{0}) - f(\alpha, t, \omega) \right\}.
\end{align}
It is straightforward to verify that turning off the sources ($\alpha = \alpha^{*}=0$) recovers the simple harmonic oscillator propagator (up to the zero-point energy phase shift).

To compute the star-exponential, we first expand the propagator. Given the properties of Grassmann integration over the auxiliary variables $\Psi'$ and $\Pi'$, we only retain the terms that contribute non-trivially to the integral. Specifically, we expand the exponential $K$ as follows
\begin{align}
    K &= 1+\pi_{f}e^{-i\omega t}\psi_{0} - \frac{1}{\omega}(1-e^{-i\omega t})(\alpha^{*}\pi_{f} +\alpha\psi_{0}) -\frac{|\alpha|^{2}}{2\omega}\bigg(-i\omega t + e^{-i\omega t} -1\bigg).
\end{align}
Substituting the center-of-mass variables $\pi_f = \Pi + \Pi'$ and $\psi_0 = \Psi - \Psi'$, we group the terms by their dependence on the primed variables
\begin{align}
    K &= 1+(\Pi +\Pi^{'})e^{-i\omega t}(\Psi-\Psi^{'}) + \frac{1}{\omega}(1-e^{-i\omega t})\big[(\Pi + \Pi^{'})\alpha^{*} +(\Psi-\Psi^{'})\alpha\big] \nonumber \\
      &\quad -\frac{|\alpha|^{2}}{2\omega}\bigg(-i\omega t + e^{-i\omega t} -1\bigg) \nonumber \\
      &= 1 -\frac{|\alpha|^{2}}{2\omega}\bigg(-i\omega t + e^{-i\omega t} -1\bigg) + \Pi\Psi e^{-i\omega t}-\Pi\Psi' e^{-i\omega t}+\Pi'\Psi e^{-i\omega t} -\Pi'\Psi' e^{-i\omega t} \nonumber \\
      &\quad +\bigg(\Pi \alpha^{*} + \Pi'\alpha^{*}+\Psi\alpha-\Psi'\alpha\bigg)\frac{1}{\omega}\bigg(1-e^{-i\omega t}\bigg).
\end{align}
Now, we multiply the propagator by the integral kernel, which can be exactly expanded due to the nilpotency of Grassmann variables as $\exp\{-2i\Pi'\Psi'/\hbar\} = 1 - \frac{2i}{\hbar}\Pi'\Psi'$, and compute:
\begin{align}
\exp\left\{-\frac{2i}{\hbar}\Pi'\Psi'\right\}K
 &= \left(1-\frac{2i}{\hbar}\Pi'\Psi'\right) \bigg[1 -\frac{|\alpha|^{2}}{2\omega}\bigg(-i\omega t + e^{-i\omega t} -1\bigg) + \Pi\Psi e^{-i\omega t} \nonumber \\
  &\quad -\Pi\Psi' e^{-i\omega t} +\Pi'\Psi e^{-i\omega t} -\Pi'\Psi' e^{-i\omega t} \nonumber \\
    &\quad +\bigg(\Pi \alpha^{*} + \Pi'\alpha^{*}+\Psi\alpha-\Psi'\alpha\bigg)\frac{1}{\omega}\bigg(1-e^{-i\omega t}\bigg)\bigg] \nonumber \\     
&= \left[ \text{Terms independent of } \Pi'\Psi' \right] \nonumber \\
    &- \frac{2i}{\hbar}\Pi' \Psi' \bigg[1 -\frac{|\alpha|^{2}}{2\omega}\bigg(-i\omega t + e^{-i\omega t} 
-1\bigg) + \Pi \Psi e^{-i\omega t} \nonumber \\
    &+ \bigg(\Pi\alpha^{*}+\Psi\alpha \bigg)\frac{1}{\omega}\bigg(1-e^{-i\omega t}\bigg) \bigg] - \Pi'\Psi' e^{-i\omega t}.
\end{align}
We now carry out the integration $\int \dots D\Psi'D\Pi'$. Recalling that only the coefficients of $\Pi'\Psi'$ survive, we obtain
\begin{align}
    \int \exp\left\{-\frac{2i}{\hbar}\Pi'\Psi'\right\}K D\Psi'D\Pi' &= -\exp\{-i\omega t\} - \frac{2i}{\hbar}\Biggl[1-\frac{|\alpha|^{2}}{2\omega}\bigg(-i\omega t + e^{-i\omega t} -1\bigg) \nonumber \\
    &\quad + \Pi\Psi\exp\{-i\omega t\} +\bigg(\Pi\alpha^{*}+\Psi\alpha\bigg)\frac{1}{\omega}\bigg(1-\exp\{-i\omega t\}\bigg)\Biggr].
\end{align}

Finally, multiplying by the normalization constant $C$ and accounting for the Grassmann nature of the exponential pre-factor (which only acts non-trivially on the scalar terms), we arrive at the closed form for the star-exponential
\begin{align}\label{star exp driv}
\mathrm{Exp}_{\star}\left(-\frac{it}{\hbar}H(\Psi,\Pi)\right) &= C\exp\left\{\frac{i}{\hbar}\Pi\Psi\right\}\int \exp\left\{-\frac{2i}{\hbar}\Pi'\Psi'\right\}K D\Psi'D\Pi' \nonumber \\
&= -\left(\frac{\hbar^{2}}{5}\right)\Biggl\{ \exp\left\{\frac{i}{\hbar}\Pi\Psi\right\}\biggl[\exp\{-i\omega t\} \nonumber \\
&\quad + \frac{2i}{\hbar}\left( 1-\frac{|\alpha|^{2}}{2\omega}\bigg(-i\omega t + e^{-i\omega t} -1) \bigg)\right) \biggr] \nonumber \\
&\quad + \frac{2i}{\hbar} \left[ \Pi\Psi\exp\{-i\omega t\} + (\Pi\alpha^{*}+\Psi\alpha)\frac{(1-\exp\{-i\omega t\})}{\omega} \right] \Biggr\}.
\end{align}

It is straightforward to verify that in the limit where the driving sources vanish ($\alpha = \alpha^{*}=0$), we recover the harmonic oscillator result (\ref{expstar}) derived in the previous section (up to the global phase shift $\omega/2$). Once again, the star-exponential obtained here does not rely on any of the standard convergence issues emerging in the standard definition as a power series of star-products.

\vskip 2truecm
\section{Feynman-Kac Formula and Deformation Quantization for Fermions}
The Feynman-Kac formula provides a powerful link between the asymptotic behavior of the system's propagator and its ground state energy \cite{Glimm:1987ylb}. Given the established connection between propagators and star-exponentials detailed in the previous section, and building upon works such as \cite{Berra-Montiel:2024ubb, Sharan:1979ej,Berra-Montiel:2020daw}, we will now establish an analogous Feynman-Kac formula for fermionic variables within the framework of deformation quantization. For a comprehensive treatment on this topic for the bosonic case, the reader is referred to Ref. \cite{Berra-Montiel:2025iuy}.

As it might be anticipated, subtle differences arise in comparison to the bosonic case, particularly concerning the different nature of the quantities defined in their respective phase spaces. In particular, we must extend the spectral decomposition analysis to the fermionic domain. The eigenvalues of the Hamiltonian operator $\widehat{H}=Q_{W}(H)$ correspond to the physical spectrum; however, the properties of the diagonal Wigner functions and the integration over Grassmann variables require a careful treatment, as we describe below.

Consider the density matrix for a pure state $\widehat{\rho}= |\psi\rangle \langle \psi|$. In the Weyl-Wigner-Groenewold-Moyal formalism, its analogous normalized Wigner function is given by
\begin{align}
    \rho_{W}(\psi,\pi) &= \hbar^{-n}\int \exp\bigg\{-i\sum_{j=1}^{n}\pi_{j}\lambda_{j}\bigg\}\bra{\psi + \frac{\hbar\lambda}{2}}\widehat{\rho}\ket{\psi - \frac{\hbar\lambda}{2}} D\lambda \nonumber\\
    &= \hbar^{-n}\int \exp\bigg\{-i\sum_{j=1}^{n}\pi_{j}\lambda_{j}\bigg\}\phi_{n}\bigg(\psi- \frac{\hbar\lambda}{2}\bigg)\overline{\phi}_n\bigg(\psi + \frac{\hbar\lambda}{2}\bigg) D\lambda.
\end{align}
Recall that the integral of the star-exponential over the phase space corresponds to the trace of the evolution operator. Utilizing the spectral decomposition of the evolution operator in the energy eigenbasis, we obtain
\begin{align}
\operatorname{Tr}\left( \exp\left\{ -\frac{it}{\hbar}\widehat{H} \right\} \right) &= \frac{1}{2\pi\hbar} \int {\rm Exp}_{\star}\left\{-{ it \over \hbar}H(\psi,\pi)\right\} D\pi D\psi \nonumber \\
    &= \int \sum_{n=0}^{\infty} e^{-i\frac{tE_n}{\hbar}} \rho_W^{(n)}(\psi, \pi)D\pi D\psi \nonumber \\
    &= \sum_{n=0}^{\infty} e^{-i\frac{tE_n}{\hbar}}\underbrace{\left( \int \rho_W^{(n)}(\psi, \pi) \, D\pi D\psi \right)}_{1} \nonumber \\
    &= \sum_{n=0}^{\infty} e^{-i\frac{tE_n}{\hbar}},
\end{align}
where $1/(2\pi\hbar)$ represents the appropriate trace normalization factor for the fermionic phase space. From this expression, performing a Wick rotation $t \to -i\tau$ leads to the statistical partition function sum. In the limit of large Euclidean time $\tau \to \infty$, the sum is dominated by the lowest energy term (ground state). Thus, we can extract the ground state energy using the following formula
\begin{align}
    E_{0} = - \lim_{\tau \to \infty } \frac{\hbar}{\tau} \ln \left( \frac{1}{2\pi\hbar} \int {\rm Exp}_{\star}\left\{-\frac{\tau H(\psi,\pi)}{\hbar}\right\}D\pi D\psi \right) \,.
\label{general FK}
\end{align}
This relation stands for the Feynman-Kac formula for the fermionic case. In the following subsections we will apply this formula for the previously analysed fermionic oscillators in order to explicitly construct their respective energy ground states.

\vskip 1truecm
\subsection{Feynman-Kac formula for the Fermi Oscillator}
Considering the star-exponential expression obtained previously for the Fermi harmonic oscillator, and evaluating the integral over the phase space, we find
\begin{align}
\int {\rm Exp}_{\star}\left\{-\frac{it}{\hbar} H(\Psi,\Pi)\right\}D\Pi D\Psi
&= \int - \left(\frac{\hbar^{2}}{5}\right) \Big[
\exp\left\{\frac{i}{\hbar}\Pi\Psi\right\}
\left(\exp\left\{-\frac{i\omega t}{2}\right\}
+\frac{2i}{\hbar}\exp\left\{i\frac{\omega t}{2}\right\}\right)  \nonumber \\
&\qquad +\frac{2i}{\hbar}\Pi\Psi
\exp\left\{-\frac{i\omega t}{2}\right\}
\Big] D\Pi D\Psi  \\ \nonumber
&= \frac{2}{5}\exp\left\{i\frac{\omega t}{2}\right\}
-\frac{3i\hbar}{5}\exp\left\{-\frac{i\omega t}{2}\right\}.
\end{align}
This implies that the ground state energy is written as
\begin{align}
E_{0} &= - \lim_{\tau\rightarrow\infty} 
\frac{\hbar}{\tau}
\ln\left| 
\frac{1}{2\pi\hbar}
\int 
{\rm Exp}_{\star}\left\{
-\frac{\tau H(\Psi,\Pi)}{\hbar}
\right\}
D\Pi\, D\Psi
\right|
\nonumber\\
&= - \lim_{\tau\rightarrow\infty} 
\frac{\hbar}{\tau}
\ln \left| 
a \exp\left\{\frac{\tau\omega}{2}\right\} + b \exp\left\{-\frac{\tau\omega}{2}\right\}
 \right| \,,
\end{align}
where, for simplicity, we have written the complex constants $a=2/5$ and $b=-3i\hbar/5$. Now, we will study the limit, showing that the nature of $a,b$ will be irrelevant at $\tau \to \infty$
\begin{align}
    \ln|ae^{\frac{\tau\omega}{2}}+be^{\frac{-\tau\omega}{2}}|
= \ln|e^{\frac{\tau\omega}{2}}(a+be^{-\tau\omega})| , \nonumber\\
    = \ln|e^{\frac{\tau\omega}{2}}| + \ln|a+be^{-\tau\omega}|, \nonumber \\
    = \frac{\tau\omega}{2} + \ln|a+be^{-\tau\omega}|.
\end{align}
Returning to the previous calculation we find
\begin{align}
E_{0} =
    - \lim_{\tau\rightarrow\infty} 
\frac{\hbar}{\tau}\bigg(
\frac{\tau\omega}{2} + \underbrace{\ln|a+be^{-\tau\omega}|}_{\approx \ln|a| \quad as\quad \tau\rightarrow\infty}\bigg) \nonumber\\
= -\frac{\hbar \omega}{2} \,.
\label{FK-fermi}
\end{align}
Since the logarithm of the constant term vanishes compared to the linear term $\tau\omega/2$ as $\tau$ tends to infinity, this confirms the exact result for the ground state energy for a fermionic system with a single degree of freedom.

\vskip 1truecm
\subsection{Feynman-Kac formula for the Driven Fermi Oscillator}
\label{FKdrivenFermi}
In an analogous manner, now we proceed to study the Feynman-Kac formula for the driven Fermi oscillator. Using (\ref{star exp driv}) we get
$$
\int {\rm Exp}_{\star}\left(-\frac{it}{\hbar}H(\Psi,\Pi\right)D\Pi D\Psi
$$
\begin{align}
= -\bigg(\frac{\hbar^{2}}{5}\bigg)\Biggl\{\bigg(\frac{i}{\hbar}\bigg)\left[\exp\{-i\omega t\}+\frac{2i}{\hbar}\bigg(1-\frac{|\alpha|^{2}}{2\omega}(-i\omega t + \exp\{-i\omega t\} -1)\bigg)\right] \nonumber \\
\quad +\frac{2i}{\hbar}\exp\{-i\omega t\}\Biggl\}.    
\end{align}
This implies the following
\begin{align}
E_{0} &= - \lim_{\tau\rightarrow\infty} 
\frac{\hbar}{\tau}
\ln\left| 
\frac{1}{2\pi\hbar}
\int 
{\rm Exp}_{\star}\left\{
-\frac{\tau H(\Psi,\Pi)}{\hbar}
\right\}
D\Pi\, D\Psi
\right| \nonumber\\
&:= - \lim_{\tau\rightarrow\infty} 
\frac{\hbar}{\tau}
\ln\left| a e^{-\tau \omega} + b\tau + c \right| \,.
\label{FK-dfermi}
\end{align}
Here the Gaussian integration over $\Psi$ and $\Pi$ has been performed, 
the linear terms and the exponential in $\tau$ have been grouped, and 
the following complex-valued constants have been introduced:
\begin{equation*}
a = \frac{1}{\pi\hbar} + \frac{i|\alpha|^{2}}{2\pi\omega\hbar^{2}}, \ \ \ b = -\frac{i|\alpha|^{2}}{\hbar}\omega, \ \ \ 
c = -i\left( \frac{|\alpha|^{2}}{\hbar\omega} - \frac{2}{\hbar} \right) \,.
\end{equation*}
We must carefully analyze the limit given the complex nature of these constants.

\vskip .5truecm

Thus the following limit must be analyzed in detail
\[
L = \lim_{\tau \to \infty} \frac{1}{\tau}\ln\bigg|a e^{-\tau \omega} + b\tau + c\bigg|,
\]
As we will see, the value of the limit will depend only on the parameter $\omega$, and not on the specific complex values of the constants $a$, $b$, and $c$. To show this, we proceed as follows.

Let $Z(\tau) = a e^{-\tau \omega} + b\tau + c$. Since $a, b, c$ are complex, $Z(\tau)$ is a complex number for any given $\tau$. The natural logarithm of a complex number $Z$ is defined as $\ln(Z) = \ln|Z| + i \cdot \text{arg}(Z)$, where $|Z|$ stands for the magnitude of $Z$, while $\text{arg}(Z)$ corresponds to its principal argument (phase). In consequence, substituting this into our limit expression, we can split the limit into its real and imaginary parts
\begin{align*}
L = \lim_{\tau \to \infty} \frac{1}{\tau} \left( \ln|Z(\tau)| + i \cdot \text{arg}(Z(\tau)) \right)    ,\\
= \underbrace{\lim_{\tau \to \infty} \frac{1}{\tau}\ln|Z(\tau)|}_{\text{Real Part}} + i \underbrace{\lim_{\tau \to \infty} \frac{1}{\tau}\text{arg}(Z(\tau))}_{\text{Imaginary Part}}.
\end{align*}

To evaluate this, we must identify the term in $Z(\tau)$ that has the largest magnitude as $\tau \to \infty$, that is, its dominant term. For very large $\tau$, the argument of $Z(\tau)$ will approach the constant argument of its dominant term, namely, $\text{arg}(Z(\tau)) \to \phi_{\text{dominant}}$. The imaginary part of the limit then becomes
\[
i \lim_{\tau \to \infty} \frac{1 \cdot \phi_{\text{dominant}}}{\tau} = 0.
\]
Since the numerator is a constant while the denominator goes to infinity, the imaginary part of the limit is always zero. Therefore, the final result is purely real and is determined entirely by the limit of the magnitude term
\[
L = \lim_{\tau \to \infty} \frac{1}{\tau}\ln|Z(\tau)|.
\]
This simplifies the problem to analysing the magnitude of the dominant term in $Z(\tau)$, which depends on the sign of $\omega$. Now, we analyse the three possible cases for the real parameter $\omega$.

\noindent
{\it Case 1: $\omega > 0$}

In this case, the term $e^{-\tau\omega}$ is a decaying exponential. As $\tau \to \infty$, the magnitudes of the terms in $Z(\tau)$ behave as follows
\begin{itemize}
    \item $|a e^{-\tau\omega}| = |a|e^{-\tau\omega} \to 0$,
    \item $|b\tau| = |b|\tau \to \infty$,
    \item $|c|$ is constant.
\end{itemize}
The dominant term is $b\tau$. Therefore, for large $\tau$, the magnitude $|Z(\tau)|$ is approximately $|b|\tau$. The limit thus becomes
\begin{align*}
    L = \lim_{\tau \to \infty} \frac{1}{\tau}\ln(|b|\tau) = \lim_{\tau \to \infty} \frac{1(\ln|b| + \ln\tau)}{\tau}.
\end{align*}

Since the linear function $\tau$ grows faster than the logarithmic function $\ln\tau$, this limit vanishes. In consequence, for $\omega>0$ we have the ground state energy $E_0=0$.

\noindent
{\it Case 2: $\omega = 0$}

If $\omega = 0$, the constant $b$ is identically zero, $b =0$. In consequence, the expression for $Z(\tau)$ simplifies to a constant, independent of $\tau$, namely, $Z(\tau) = a+c$. The limit in this case is therefore
\[
L = \lim_{\tau \to \infty} \frac{\ln(a+c)}{\tau}.
\]
Since $\ln(a+c)$ is just a complex constant, the limit also vanishes. As for the case $\omega>0$, we recover a ground state energy $E_0=0$.

\noindent
{\it Case 3: $\omega < 0$}

In this case, it is helpful to write $\omega = -|\omega|$. The exponential term becomes $e^{-\tau\omega} = e^{\tau|\omega|}$, which is a growing exponential. The magnitudes behave as
\begin{itemize}
    \item $|a e^{\tau|\omega|}| \to \infty$ (exponential growth),
    \item $|b\tau| \to \infty$ (linear growth),
    \item $|c|$ is constant.
\end{itemize}
Exponential growth is much faster than linear growth, so the dominant term is $a e^{\tau|\omega|}$. The magnitude $|Z(\tau)|$ is approximately $|a|e^{\tau|\omega|}$. The limit becomes
$$
L = \lim_{\tau \to \infty} \frac{1}{\tau}\ln(|a|e^{\tau|\omega|})
= \lim_{\tau \to \infty} \frac{1}{\tau}(\ln|a| + \ln(e^{\tau|\omega|}))
$$
$$
= \lim_{\tau \to \infty} \frac{1}{\tau}(\ln|a| + \tau|\omega|) 
= \lim_{\tau \to \infty} \left( \frac{\ln|a|}{\tau} + \frac{\tau|\omega|}{\tau} \right).
$$
The first term goes to zero, thus leaving
\[
L = |\omega|.
\]
Since $\omega < 0$, we have $|\omega| = -\omega$. Thus, the value of the ground state energy $E_0$ equals $ \omega$, where we have considered the overall minus sign in formula (\ref{general FK}).

As discussed above, the final value of the limit depends solely on the sign of $\omega$. The results for the distinct values of $\omega$ are summarized below.

\begin{table}[htbp]
\centering
\caption{Final value of the limit $L$ based on the condition on $\omega$.}
\vskip .3truecm
\label{tab:limit-conditions}
\begin{tabular}{|l|c|}
\hline
\textbf{Condition on $\omega$} & \textbf{Final Value of the Limit ($L$)} \\
\hline
$\omega > 0$ & $0$ \\
\hline
$\omega = 0$ & $0$ \\
\hline
$\omega < 0$ & $-\omega$ \\
\hline
\end{tabular}
\end{table}

If we then recall that the expression for $E_{0}$ has a global minus sign,  modulo $\hbar$, we find
\begin{align}
    E_{0} = \begin{cases} 
      0 & \omega\geq 0 \\
      \omega & \omega<0
   \end{cases}
\end{align}
Now, if we recall the result for the eigenenergies determined for the Hamiltonian we have
\begin{eqnarray}
    \lambda &=& 
    \frac{\omega \ \pm\ \sqrt{\omega^{2}+ 4|\alpha|^{2}}}{2}   = 
    \frac{\omega}{2} \ \pm\ \frac{1}{2}\sqrt{\omega^{2}+4|\alpha|^{2}} \nonumber \\
    &\approx& 
    \begin{cases}
        \omega   \\
        0
    \end{cases}
     \quad \iff |\alpha| \ll |\omega|  \,. \label{approx-eigen}
\end{eqnarray}
In Appendix A, we study in greater detail some aspects in order to validate the approximation considered in (\ref{approx-eigen}), as well as the different physical conditions (labelled by $\omega$) considered in Table 2 below, where we have considered $g:=|\alpha|$.
\medskip

\begin{table}[H]
\centering
\caption{Approximate eigenvalues $\lambda_\pm$ under different physical conditions.}
\vskip .3truecm
\label{tab:eigenvalues}
\begin{tabular}{|l|l|l|}
\hline
\textbf{Region} & \textbf{Condition} & \textbf{Approximate eigenvalues } $\lambda_\pm$ \\ \hline
(1) Resonant & $\omega \to 0$, $g$ not weak &
$\displaystyle
\lambda_\pm \approx \pm g + \frac{\omega}{2} \pm \frac{\omega^{2}}{8g}
\quad\big(O\!\left((\omega/g)^3\right)\big)
$
\\ \hline
(2) Dispersive ($\omega>0$) & $|\omega|\gg g$ &
$\displaystyle
\lambda_+ \approx \omega + \frac{g^{2}}{\omega},
\qquad
\lambda_- \approx -\frac{g^{2}}{\omega}
\quad\big(O(g^4/\omega^3)\big)
$
\\ \hline
(2) Dispersive ($\omega<0$) & $|\omega|\gg g$ &
$\displaystyle
\lambda_- \approx \omega + \frac{g^{2}}{\omega},
\qquad
\lambda_+ \approx -\frac{g^{2}}{\omega}
\quad\big(O(g^4/\omega^3)\big)
$
\\ \hline
(3) Strong coupling & $g \gg |\omega|$ &
$\displaystyle
\lambda_\pm \approx \pm g + \frac{\omega}{2} \pm \frac{\omega^{2}}{8g}
\quad\big(O\!\left((\omega/g)^3\right)\big)
$
\\ \hline
(4) Weak coupling & $g \to 0$ &
$\displaystyle
\lambda_+ \to \omega,
\qquad
\lambda_- \to 0
$
\\ \hline
\end{tabular}
\end{table}

\par As shown in Table 2, our exact calculation is consistent with the standard approximations in the weak coupling regime.

\vskip 1truecm
\section{Final Remarks}
\label{S-FinalRemarks}
In this work, we have successfully extended to the realm of fermionic systems the correspondence between the star-exponentials, naturally emerging within the context of DQ, and the path integral propagators commonly associated with the quantum field theory perspective. We developed a consistent formalism that yields an exact expression for the fermionic evolution operator symbol straightforwardly obtained by means of the quantum propagator.

As a primary application, we derived the fermionic analog of the Feynman-Kac formula, providing a robust method for extracting ground state energies within the deformation quantization framework. Additionally, the derivation of a general fermionic Feynman-Kac formula (\ref{general FK}) provides an alternative computational pathway for extracting ground state energies directly from phase-space expressions.

The formalism was validated by analysing two fundamental fermionic examples: the simple harmonic oscillator (\ref{FK-fermi}) and the driven fermionic oscillator (\ref{FK-dfermi}). We obtain an explicit relation (\ref{exponential-star-general}) which connects the star-exponential function with the path integral propagator of a generic fermionic system. We also found an explicit integral representation for the determination of the star-exponential for the cases of the Fermi oscillator and the driven Fermi oscillator in expressions (\ref{expstar}) and (\ref{star exp driv}) respectively. For such mechanical systems, several technical challenges were addressed. First, we derived a concise exact expression for the fermionic propagator using the coherent state path integral. Second, we resolved the ambiguities regarding boundary conditions and Grassmann parity preservation. While initial heuristic approaches suggested the need for auxiliary constraints on the phase space integration, our results demonstrate that a rigorous application of the star-product trace definition naturally handles these subtleties.

Looking forward, a natural extension of this research may involve the reasonable unification of bosonic and fermionic degrees of freedom. This will lead towards a supersymmetric (SUSY) quantum mechanics implementation, where the interplay between the two sectors could further elucidate the geometric structure of quantization. A useful guide for this endeavour would be the formalism established in \cite{Galaviz:2006ni} combined with the standard path integral description for SUSY systems. Furthermore, the present work provides a preliminary step for investigating more general contexts, including quantum field theory \cite{Mrowczynski:1987dm,Mrowczynski:1994nf,Garcia-Compean:1999dtf,Galaviz:2007te,CarballoPerez:2011ig,Polyakov:2016qqt,Mueller:2019gjj} and, potentially, the string theory setup \cite{Garcia-Compean:2000iju}, where deformation quantization continues to offer significant challenges.

\break
\centerline{\bf Acknowledgements} \vspace{.5cm} 
A. Kafuri (CVU 1314608) acknowledges the financial support from the Secretaría de Ciencia, Humanidades, Tecnología e Innovación (SECIHTI) through the National Scholarship program. AM acknowledges financial support from COPOCYT under project 2467 HCDC/2024/SE-02/16 (Convocatoria 2024-03, Fideicomiso 23871). JBM acknowledges financial support from the Marcos Moshinsky Foundation.

\vskip 1truecm
\appendix

\section{Analysis of approximation regions: parameter space}

We will now provide the details on the analysis for each region summarized in Table 2 of section \ref{FKdrivenFermi}. As before, we consider here $g:=|\alpha|$.

\subsection{Case 1: Resonant and Strong Coupling ($|\omega/g| \ll 1$)}
This condition applies to both the \textit{Resonant} region ($\omega \to 0$, $g$ not weak) and the \textit{strong coupling} region ($g \gg |\omega|$), as both are mathematically controlled by the dominance of $g$ over $\omega$. We start by factoring out the dominant term, $2g$, from the square root in the eigenvalue formula
    \begin{align*}
        \lambda_\pm = \frac{\omega \pm \sqrt{4g^2 + \omega^2}}{2} = \frac{\omega \pm \sqrt{4g^2 \left(1 + \frac{\omega^2}{4g^2}\right)}}{2} = \frac{\omega \pm 2g \sqrt{1 + \left(\frac{\omega}{2g}\right)^2}}{2}.
\end{align*}
and apply the Taylor series expansion for $\sqrt{1+x}$, obtaining the approximate result
   \begin{align*}
        \lambda_\pm &\approx \frac{\omega \pm 2g \left(1 + \frac{\omega^2}{8g^2}\right)}{2} = \frac{\omega}{2} \pm \left(g + \frac{\omega^2}{4g}\right).
\end{align*}
Finally, rearranging the terms gives the desired approximation for the cases $|\omega/g|\ll 1$
    \[
    \lambda_\pm \approx \pm g + \frac{\omega}{2} \pm \frac{\omega^2}{8g}.
\]

\subsection{Case 2: Dispersive ($\omega > 0$ and $|\omega| \gg g$)}
This condition corresponds to the limit where $|g/\omega| \ll 1$. Since $\omega$ is the dominant term in such cases, and given that $\omega$ is positive, we factor it out from the square root in the expression for $\lambda_\pm$. For $\omega>0$, $\sqrt{\omega^2}=\omega$, and thus
    \begin{align*}
        \lambda_\pm = \frac{\omega \pm \sqrt{\omega^2 \left(1 + \frac{4g^2}{\omega^2}\right)}}{2} = \frac{\omega \pm \omega \sqrt{1 + \frac{4g^2}{\omega^2}}}{2}.
\end{align*}

Once again, Taylor expanding the square root term in the last expression we get the approximate eigenvalues
    \begin{eqnarray}
    \lambda_+ &\approx & \frac{\omega + \omega\left(1 + \frac{2g^2}{\omega^2}\right)}{2} = \frac{2\omega + \frac{2g^2}{\omega}}{2} = {\omega + \frac{g^2}{\omega}} \,,
\nonumber\\
	\lambda_- &\approx & \frac{\omega - \omega\left(1 + \frac{2g^2}{\omega^2}\right)}{2} = \frac{-\frac{2g^2}{\omega}}{2} = {-\frac{g^2}{\omega}}  \,.
\nonumber
    \end{eqnarray}

\subsection{Case 3: Dispersive ($\omega < 0$ and $|\omega| \gg g$)}
The condition is still $|g/\omega| \ll 1$, as in the previous case but the difference now is that $\omega$ is negative. For this case, we proceed as before. We start by factoring out the $|\omega|$ from the square root in the eigenvalue expression for $\lambda_\pm$. It is critical to recognize that since $\omega < 0$, we have $\sqrt{\omega^2} = |\omega| = -\omega$
    \begin{align*}
        \lambda_\pm = \frac{\omega \pm \sqrt{\omega^2 \left(1 + \frac{4g^2}{\omega^2}\right)}}{2} = \frac{\omega \pm (-\omega) \sqrt{1 + \frac{4g^2}{\omega^2}}}{2} \,.
\end{align*}
Taylor expanding the square root as in the previous cases we get $\sqrt{1 + 4g^2/\omega^2} \approx 1 + 2g^2/\omega^2$, and thus we find the approximate eigenvalues
        \begin{eqnarray}
        \lambda_+ &\approx & \frac{\omega - \omega\left(1 + \frac{2g^2}{\omega^2}\right)}{2} = \frac{\omega - \omega - \frac{2g^2}{\omega}}{2} = {-\frac{g^2}{\omega}} \,, \nonumber\\
         \lambda_- &\approx & \frac{\omega + \omega\left(1 + \frac{2g^2}{\omega^2}\right)}{2} = \frac{2\omega + \frac{2g^2}{\omega}}{2} = {\omega + \frac{g^2}{\omega}} \,.
\nonumber
        \end{eqnarray}

\subsection{Case 4: Weak Coupling ($g \to 0$)}
This condition implies that $g = |\alpha| \to 0$.

In this case, we take the limit of the exact formula for $\lambda_\pm$ as $g \to 0$
    \begin{align*}
        \lim_{g \to 0} \lambda_\pm = \lim_{g \to 0} \frac{\omega \pm \sqrt{\omega^2 + 4g^2}}{2} = \frac{\omega \pm \sqrt{\omega^2}}{2} = \frac{\omega \pm |\omega|}{2} \,.
\end{align*}
The result depends on the sign of $\omega$. If ${\omega > 0}$, then $|\omega|=\omega$ and thus the corresponding limits are $\lambda_+ \to \omega$ and $\lambda_- \to 0$. On the contrary, if ${\omega < 0}$, those limits are interchanged, namely, $\lambda_+ \to 0$ and $\lambda_- \to \omega$.
\newpage


\begin{thebibliography}{99}

\bibitem{Bayen:1977ha}
F.~Bayen, M.~Flato, C.~Fronsdal, A.~Lichnerowicz and D.~Sternheimer,
``Deformation theory and quantization. 1. Deformations of symplectic structures,''
Ann.~Phys. \textbf{111}, 61 (1978)
doi:10.1016/0003-4916(78)90224-5

\bibitem{Bayen:1977hb}
F.~Bayen, M.~Flato, C.~Fronsdal, A.~Lichnerowicz and D.~Sternheimer,
``Deformation theory and quantization. 2. Physical applications,''
Ann.~Phys. \textbf{111}, 111 (1978)
doi:10.1016/0003-4916(78)90225-7

\bibitem{Wilde-Lecomte}
 M. De Wilde and P.B.A. Lecomte,
``Existence of star-products and of formal deformations of the Poisson Lie algebra of 
arbitrary symplectic manifolds,''
Lett. Math. Phys. \textbf{7}, 487 (1983)
https://doi.org/10.1007/BF00402248

\bibitem{Omori-1991}
 H. Omori, Y. Maeda and A. Yoshioka,
``Weyl manifolds and deformation quantization,''
Adv. Math. \textbf{85}, 224 (1991)
https://doi.org/10.1016/0001-8708(91)90057-E

\bibitem{Fedosov:1994zz}
B.~Fedosov,
``A simple geometrical construction of deformation quantization,''
J. Diff. Geom. \textbf{40}, no.2, 213-238 (1994)

\bibitem{Fedosov:1996fu}
B.~Fedosov,
``Deformation quantization and index theory,''  (Akademie Verlag, Berlin, 1996).

\bibitem{Kontsevich:1997vb}
M.~Kontsevich,
``Deformation quantization of Poisson manifolds. 1.,''
Lett. Math. Phys. \textbf{66}, 157-216 (2003)
doi:10.1023/B:MATH.0000027508.00421.bf
[arXiv:q-alg/9709040 [math.QA]].
	
\bibitem{Todorov:2012gy}
I.~Todorov,
``Quantization is a mystery,''
Bulg. J. Phys. \textbf{39}, 107-149 (2012)
[arXiv:1206.3116 [math-ph]].

\bibitem{Curtright:2014sli}
T.~L.~Curtright, D.~B.~Fairlie and C.~K.~Zachos,
``A concise treatise on quantum mechanics in phase space,''
(World Scientific Publishing Co Pte Ltd, 2014),
ISBN 978-981-4520-45-4
doi:10.1142/8870

\bibitem{Hirshfeld:2002yjb}
A.~C.~Hirshfeld and P.~Henselder,
``Deformation quantization in the teaching of quantum mechanics,''
Am. J. Phys. \textbf{70}, 537-547 (2002)
doi:10.1119/1.1450573
[arXiv:quant-ph/0208163 [quant-ph]].

\bibitem{feynman2010quantum}
R.~P.~Feynman, A.~R.~Hibbs and D.~F.~Styer,
``Quantum mechanics and path integrals: Emended edition,''
(Dover Publ., 2010),
ISBN: 978-0-486-47722-0

\bibitem{Weinberg:1995mt}
S.~Weinberg,
``The quantum theory of fields. Vol. 1: Foundations,''
(Cambridge University Press, 2005),
ISBN 978-0-521-67053-1, 978-0-511-25204-4
doi:10.1017/CBO9781139644167

\bibitem{das2019field}
A.~Das,
``Field theory: A path integral approach,''
World Sci. Lect. Notes Phys. \textbf{83}, 1-456 (2019)
doi:10.1142/11369

\bibitem{Zinn-Justin-book-2004}
J.~Zinn-Justin,
``Path integrals in quantum mechanics,''
(Oxford University Press, 2004), https://doi.org/10.1093/acprof:oso/9780198566748.001.0001,
Online ISBN: 9780191717994, Print ISBN: 9780198566748

\bibitem{Dito:2005xt}
G.~Dito and F.~J.~Turrubiates,
``The damped harmonic oscillator in deformation quantization,''
Phys. Lett. A \textbf{352}, 309-316 (2006)
doi:10.1016/j.physleta.2005.12.013
[arXiv:quant-ph/0510150 [quant-ph]].

\bibitem{cadavid-1991}
A. C. Cadavid and  M. Nakashima,
``The star-exponential and path integrals on compact groups,''
Lett. Math. Phys. \textbf{23}, 111-115 (1991).

\bibitem{Omori-2004}
H. Omori, Y. Maeda, N. Miyazaki, A. Yoshioka,
``Star-exponential functions as two-valued elements,''
In: Marsden, J., Ratiu, T. (eds.) The breadth of symplectic
and Poisson geometry, Progress in Mathematics, vol.
232, pp. 483-492,
Birkh¨auser, Boston (2004), arXiv:0711.3668. 

\bibitem{Omori:2012xv}
H.~Omori, Y.~Maeda, N.~Miyazaki and A.~Yoshioka,
``Deformation expression for elements of algebras (VII) --Vacuum/Pseudo-vacuum representations--,''
AIP Conf. Proc. \textbf{1564}, 184 (2013).

\bibitem{Bieliavsky:2013eca}
P. Bieliavsky, A. de Goursac and F. Spinnler, ``Non-formal deformation quantization and star-exponential of the Poincaré group,'' In: Kielanowski, P., Ali, S., Odesskii, A., Odzijewicz, A., Schlichenmaier, M., Voronov, T. (eds) Geometric Methods in Physics.
Trends in Mathematics. Birkhäuser, Basel (2013). $https://{\rm doi.org}/10.1007/978-3-0348-0645-9_1$

\bibitem{Yoshioka-2018}
A. Yoshioka,
``Star-products, star-exponentials, and star functions,''
In: Diagana, T., Toni, B. (eds) Mathematical Structures and Applications.
STEAM-H: Science, Technology, Engineering, Agriculture, Mathematics and
Health. Springer, Cham. doi.org/10.1007/978-3-319-97175-9 20 (2018).

\bibitem{Yoshioka:2019tjf}
A.~Yoshioka,
``Star exponentials in star product algebra,'' In: Kielanowski, P., Odzijewicz, A., Previato, E. (eds) Geometric Methods in Physics XXXVI.
Trends in Mathematics. Birkhäuser, Cham.  (2019) $https://{\rm doi.org}/10.1007/978-3-030-01156-7_17$

\bibitem{Berra-Montiel:2025vhf}
J.~Berra-Montiel, D.~Contreras-Bear, A.~Molgado and M.~Sanchez-Cordova,
``Star exponentials and Wigner functions for time-dependent harmonic oscillators,''
[arXiv:2503.06384 [quant-ph]].

\bibitem{Berra-Montiel:2024ubb}
J.~Berra-Montiel, H.~García-Compeán and A.~Molgado,
``Star-exponentials from propagators and path integrals,''
Ann.~Phys. \textbf{468}, 169744 (2024)
doi:10.1016/j.aop.2024.169744
[arXiv:2404.08815 [math-ph]].
	
\bibitem{Berra-Montiel:2025iuy}
J.~Berra-Montiel, H.~García-Compeán and A.~Molgado,
``The Feynman-Kac formula in deformation quantization,''
Eur. Phys. J. Plus \textbf{140}, no.9, 881 (2025)
doi:10.1140/epjp/s13360-025-06820-0
[arXiv:2502.03624 [math-ph]].

\bibitem{Casalbuoni:1976tz}
R.~Casalbuoni,
``The classical mechanics for Bose-Fermi systems,''
Nuovo Cim. A \textbf{33}, 389 (1976)
doi:10.1007/BF02729860

\bibitem{Casalbuoni:1975bj}
R.~Casalbuoni,
``On the quantization of systems with anticommutating variables,''
Nuovo Cim. A \textbf{33}, 115 (1976)
doi:10.1007/BF02748689

\bibitem{Berezin:1976eg}
F.~A.~Berezin and M.~S.~Marinov,
``Particle spin dynamics as the Grassmann variant of classical mechanics,''
Ann.~Phys. \textbf{104}, 336 (1977)
doi:10.1016/0003-4916(77)90335-9

\bibitem{Marnelius:1988pj}
R.~Marnelius,
``Fermionic quantum mechanics and superfields,''
Int. J. Mod. Phys. A \textbf{5}, 329 (1990)
doi:10.1142/S0217751X90000143

\bibitem{Lahiri:1990sv}
A.~Lahiri, P.~K.~Roy and B.~Bagchi,
``Supersymmetry in quantum mechanics,''
Int. J. Mod. Phys. A \textbf{5}, 1383-1456 (1990)
doi:10.1142/S0217751X90000647

\bibitem{Hirshfeld:2002ki}
A.~C.~Hirshfeld and P.~Henselder,
``Deformation quantization for systems with fermions,''
Ann.~Phys. \textbf{302}, 59-77 (2002)
doi:10.1006/aphy.2002.6302

\bibitem{Hirshfeld:2004aj}
A.~C.~Hirshfeld, P.~Henselder and T.~Spernat,
``Cliffordization, spin and fermionic star-products,''
Ann.~Phys. \textbf{314}, 75-98 (2004)
doi:10.1016/j.aop.2004.06.008
[arXiv:quant-ph/0404168 [quant-ph]].

\bibitem{Galaviz:2006ni}
I.~Galaviz, H.~García-Compeán, M.~Przanowski and F.~J.~Turrubiates,
``Weyl-Wigner-Moyal formalism for Fermi classical systems,''
Ann.~Phys. \textbf{323}, 267-290 (2008)
doi:10.1016/j.aop.2007.04.004
[arXiv:hep-th/0612245 [hep-th]].

\bibitem{Odendahl:2007gzs}
S.~Odendahl and P.~Henselder,
``Spin description in the star-product and the path integral formalism,''
Phys. Lett. A \textbf{372}, 2552 (2008)
doi:10.1016/j.physleta.2007.10.035
[arXiv:0705.3601 [quant-ph]].

\bibitem{Lin:2022nns}
B.~Lin and T.~Heng,
``Deformation quantization for systems with second-class constraints in deformed fermionic phase space,''
Mod. Phys. Lett. A \textbf{37}, no.17, 2250107 (2022)
doi:10.1142/S0217732322501073

\bibitem{Mrowczynski:2012ps}
S.~Mrowczynski,
``Wigner functional of fermionic fields,''
Phys. Rev. D \textbf{87}, no.6, 065026 (2013)
doi:10.1103/PhysRevD.87.065026
[arXiv:1212.5703 [hep-th]].

\bibitem{DeWitt:2012mdz}
B.~S.~DeWitt,
``Supermanifolds,''
Cambridge Univ. Press, 2012,
ISBN 978-1-139-24051-2, 978-0-521-42377-9
doi:10.1017/CBO9780511564000

\bibitem{Engl:2014rwj}
T.~Engl, P.~Pl{\"o}{\ss}l, J.~D.~Urbina and K.~Richter,
``The semiclassical propagator in fermionic Fock space,''
Theor. Chem. Acc. \textbf{133}, 1563 (2014)
doi:10.1007/s00214-014-1563-9
[arXiv:1409.4196 [physics.chem-ph]].

\bibitem{Kafuri:2026icq}
A.~Kafuri,
``Path integrals and deformation quantization: the fermionic case,''
MSc thesis, Cinvestav, Mexico (2025)
[arXiv:2602.00367 [quant-ph]].

\bibitem{Glimm:1987ylb}
J.~Glimm and A.~Jaffe,
``Quantum physics: A functional integral point of view,''
Springer, 1987,
ISBN 978-0-387-96477-5, 978-1-4612-4728-9
doi:10.1007/978-1-4612-4728-9

\bibitem{Sharan:1979ej}
P.~Sharan,
``$*$-Product representation of path integrals,''
Phys. Rev. D \textbf{20}, 414 (1979)
doi:10.1103/PhysRevD.20.414

\bibitem{Berra-Montiel:2020daw}
J.~Berra-Montiel,
``Star product representation of coherent state path integrals,''
Eur. Phys. J. Plus \textbf{135}, no.11, 906 (2020)
doi:10.1140/epjp/s13360-020-00930-7
[arXiv:2007.02483 [quant-ph]].

\bibitem{Mrowczynski:1987dm}
S.~Mrowczynski,
``Some remarks on the gauge covariant Wigner function of Yang-Mills fields,''
J. Phys. G \textbf{13}, L33-L37 (1987)
doi:10.1088/0305-4616/13/4/001

\bibitem{Mrowczynski:1994nf}
S.~Mrowczynski and B.~Muller,
``Wigner functional approach to quantum field dynamics,''
Phys. Rev. D \textbf{50}, 7542-7552 (1994)
doi:10.1103/PhysRevD.50.7542
[arXiv:hep-th/9405036 [hep-th]].

\bibitem{Garcia-Compean:1999dtf}
H.~García-Compeán, J.~F.~Pleba\'nski, M.~Przanowski and F.~J.~Turrubiates,
``Deformation quantization of classical fields,''
Int. J. Mod. Phys. A \textbf{16}, 2533-2558 (2001)
doi:10.1142/S0217751X01003652
[arXiv:hep-th/9909206 [hep-th]].

\bibitem{Galaviz:2007te}
I.~Galaviz, H.~García-Compeán, M.~Przanowski and F.~J.~Turrubiates,
``Deformation quantization of Fermi fields,''
Ann.~Phys. \textbf{323}, 827-844 (2008)
doi:10.1016/j.aop.2007.05.006
[arXiv:hep-th/0703125 [hep-th]].

\bibitem{CarballoPerez:2011ig}
B.~Carballo Pérez and H.~García-Compeán,
``Rarita-Schwinger quantum free field via deformation quantization,''
Found. Phys. \textbf{42}, 362-368 (2012)
doi:10.1007/s10701-011-9605-9
[arXiv:1104.5067 [hep-th]].

\bibitem{Polyakov:2016qqt}
E.~A.~Polyakov,
``Grassmann phase-space methods for fermions: uncovering classical probability structure,''
Phys. Rev. A \textbf{94}, 062104 (2016)
doi:10.1103/PhysRevA.94.062104
[arXiv:1609.06360 [quant-ph]].

\bibitem{Mueller:2019gjj}
N.~Mueller and R.~Venugopalan,
``Constructing phase space distributions with internal symmetries,''
Phys. Rev. D \textbf{99}, no.5, 056003 (2019)
doi:10.1103/PhysRevD.99.056003
[arXiv:1901.10492 [hep-th]].

\bibitem{Garcia-Compean:2000iju}
H.~García-Compeán, J.~F.~Pleba\'nski, M.~Przanowski and F.~J.~Turrubiates,
``Deformation quantization of bosonic strings,''
J. Phys. A \textbf{33}, 7935-7953 (2000)
doi:10.1088/0305-4470/33/44/305
[arXiv:hep-th/0002212 [hep-th]].

\end{thebibliography}
\end{document}